\documentclass{WileyMSP-template}
\usepackage{multicol}
\usepackage{float}
\usepackage{amsmath}
\usepackage{mathtools}
\usepackage{tikz}
\usetikzlibrary{quantikz2}
\usepackage{setspace}
\usepackage{array}
\usepackage{hyperref}

\begin{document}

\pagestyle{fancy}
\rhead{\includegraphics[width=2.5cm]{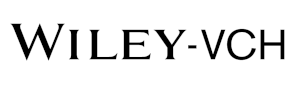}}

\title{Shock-Capturing Quantum Algorithm for the Linear Advection Operator}

\maketitle


\author{Samuel Hagele*}
\author{William Gregory}
\author{Yuan Shi}



\begin{affiliations}
Department of Physics, Center for Integrated Plasma Studies, University of Colorado Boulder, Boulder, CO 80309, USA \\
*samuel.hagele@colorado.edu
\end{affiliations}



\begin{abstract}
The linear advection operator is an ubiquitous building block in fluid and plasma problems. We develop a quantum algorithm for enacting the operator. When the advection velocity is constant in space, our algorithm is exponentially more efficient per time step than classical and avoids spurious oscillations near steep gradients. 
The algorithm is most cleanly illustrated using the one-dimensional advection equation on a uniform spatial grid with periodic boundary conditions, which can be extended to higher dimensions.
The algorithm uses a first-order upwind scheme, which captures discontinuities in the wave envelope but is not unitary. 
We embed the non-unitary upwind scheme using Linear Combinations of Unitaries (LCUs), and develop an efficient quantum gate decomposition of the upwind unitary, which performs one step of advection using $O(n^2)$ two-qubit gates, where $N=2^n$ is the number of spatial grid points, as opposed to a classical computer which costs $O(N)$. 
Although LCUs introduces a small bounded probability of failure per time step, we show that the accumulation of failures does not lead to exponential-in-time complexity as one would naively expect. 
Moreover, when LCUs fails, we develop a probabilistic scheme to recover from the failure state, which avoids a full restart of the simulation. The failure recovery scheme uses quantum Fourier transform (QFT) and effectively achieves quantum indefinite integration of an unknown quantum state. The recovery, which can itself fail, is more efficient than a full restart if the wave envelop is well-resolved to include only low Fourier modes. 
We emulate our scheme classically and demonstrate small problems on Quantinuum's trapped-ion qubits. 
Our quantum algorithm provides a subroutine for physics simulations that involve linear advection.

\end{abstract}


\section{Introduction}
The ability to solve differential equations numerically is essential for understanding many systems, including fluid and plasma systems, where the dimensionality of the discretized system is high. 
Practical applications for solving high dimensional partial differential equations (PDEs) include aircraft design \cite{gaitan_finding_2020, Ye24}, weather prediction \cite{Tennie23}, and fusion device designs \cite{Joseph23}. 
Quantum computing offers a potential reduction in complexity by encoding solutions into a quantum state whose dimension grows exponentially with the number of qubits. 
For linear ordinary differential equations (ODEs), quantum algorithms with complexity that is polylogarithmic with the inverse error have been developed \cite{berry_quantum_2017}. Subsequent work extended these ideas to linear PDEs using spectral and finite-difference methods, demonstrating polylogarithmic dependence on the number of spatial degrees of freedom under appropriate conditions \cite{childs_quantum_2020, childs_high-precision_2021}.

Nonlinear differential equations are significantly more difficult to solve on quantum computers due to the linearity of quantum mechanics. 
Multiple approaches have been explored, but attaining quantum advantages beyond polynomial speedup is uncommon. 
One approach is Carleman and Koopman-von Neumann embedding, which approximates nonlinear dynamics in a finite dimensional space by linear dynamics in an extended infinite-dimensional space \cite{engel_linear_2021, Joseph_2023, Gonzalez-Conde25}. 
This technique has been used to simulate strongly dissipative nonlinear differential equations and has recently been extended to broader classes of nonlinear problems \cite{liu_efficient_2021, wu_quantum_2025}. 
Another approach to simulate nonlinear dynamics uses hybrid classical-quantum computing, where linear steps harness speedups provided by quantum computers, while nonlinear steps are processed classically \cite{Kyriienko21, Andress25}. Using this approach, recent work has investigated direct quantum simulation of the nonlinear Vlasov–Maxwell equations \cite{higuchi_quantum_2025}. Although these approaches demonstrate the potential of quantum computing for fluid and plasma dynamics, many are designed primarily for fault-tolerant quantum computers and rely on quantum simulation primitives whose practical circuit implementations can be costly on near-term hardware \cite{NOVIKAU2025109498,NOVIKAU2026115132,PhysRevA.110.062214}.
An alternative approach that works on current or near-term quantum hardware is the second-quantization approach, where nonlinear equations are promoted to linear dynamics in Hilbert space \cite{Shi21, Shi24}. Using this approach, wave scattering has been simulated on a superconducting quantum device \cite{Sundar26}, and nonlinear Vlasov-Poisson systems may be simulated on near-term hardware platforms \cite{May2025second}.

An important step in the efficient simulation of fluid and plasma dynamics is solving the advection equation. If the advection operator can be realized on quantum hardware efficiently, it can be used as a subroutine in quantum solvers using Hamiltonian splitting, which has already been used to classically solve the Vlasov-Maxwell equations \cite{crouseilles_hamiltonian_2015}. Brearley and Laizet developed a quantum advection algorithm using Hamiltonian simulation, achieving exponential quantum speedup \cite{brearley_quantum_2024}. They use a central finite-difference first-derivative operator, which allows for the construction of a suitable Hamiltonian. However, this central difference derivative does not provide enough dissipation to be stable near discontinuities, producing unphysical oscillations around discontinuities of wave envelopes. Over et al. propose a quantum advection algorithm that explicitly defines a block encoding of the time-marching advection operator, but a central finite-difference scheme is still used, resulting in spurious oscillations near discontinuities \cite{over_quantum_2025}. In fluid and plasma problems, discontinuities often arise in the form of shock waves, which must be captured for many practical use cases. For example, in Inertial Confinement Fusion (ICF), shock-wave dynamics are essential \cite{hurricane_physics_2023}. To simulate phenomena involving discontinuities on quantum computers, shock-capturing quantum advection algorithms must be developed.

In this paper, we present a method that solves the advection equation on a quantum computer, which captures shock-like discontinuities and can be used on near-term quantum hardware with exponential speedup per time step over classical algorithms. We use an upwind numerical scheme inspired by classical methods that is numerically stable near discontinuities \cite{Courant52}. The upwind scheme is dissipative, introducing unphysical numerical diffusion that nevertheless mimics dissipation caused by viscosity and other nonconservative forces. The upwind advection operator is non-unitary, so we implement it using a Linear Combination of Unitaries (LCUs) \cite{childs_hamiltonian_2012}, introducing a small bounded probability of failure at each time step. Despite the non-zero failure probability per step, we show that the exponential quantum speedup is preserved asymptotically. Additionally, we develop a probabilistic failure recovery algorithm using an indefinite integration technique based on Quantum Fourier Transform (QFT) \cite{coppersmith_approximate_2002}, which allows a failed advection step to continue without restarting the simulation from scratch. For both LCUs advection and failure recovery algorithms, we present explicit gate-level decomposition, and we show that the gate complexity per time step is $O(n^2)$, where $N = 2^n$ is the number of spatial grid points.

In Sec.~2, we present our quantum advection algorithm, including gate decomposition. In Sec.~3, we develop a failure recovery scheme which effectively achieves quantum indefinite integration. In Sec.~4, we analyze the algorithm's complexity and probability of failure, proving that robust quantum speedup can be attained. In Sec.~5, we present demonstrations of the advection and failure recovery algorithms on classical simulators and actual quantum devices. In Sec.~6, we discuss strengths and weaknesses of our algorithms and future directions.

\section{Quantum Advection Algorithm}
The linear advection equation describes the transport of conserved quantities. Suppose $\mathbf{c}$ is the velocity of the transport, then the advection equation is $\partial_t A + \mathbf{c}\cdot\nabla A = 0$,
where $A(\mathbf{x},t)$ is the waveform of the conserved quantity. 
When $\mathbf{c}$ is a constant, the linear advection equation is analytically solvable: Suppose that $a(\mathbf{x}) = A(\mathbf{x}, t=0)$ is the initial waveform, then the solution is $A(\mathbf{x},t) = a(\mathbf{x}-\mathbf{c}t)$. 
A numerical solution to the advection equation becomes necessary when advection is only part of the problem.
For example, the Vlasov equation contains a nonlinear advection in velocity space in addition to a linear advection in position space. To solve more general problems numerically, one must first be able to solve the advection problem. 
In this paper, we focus on linear advection with constant $\mathbf{c}$. Because the problem is essentially one dimensional (1D), we choose a coordinate system such that $\mathbf{c}$ is along the $x$ direction. The 1D advection equation is
\begin{equation}
    (\partial_t + c\partial_x) A = 0,
\end{equation} 
where the scalar $c$ can be positive or negative. 
The advection equation is in fact a Schrödinger equation $i\hbar\partial_t A = \hat{H} A$, where the Hamiltonian $\hat{H}=c\hat{p}_x$ is proportional to the momentum operator $\hat{p}_x=-i\hbar\partial_x$. 
The dynamics generated by $\hat{H}$ are therefore spatial translations generated by $\hat{p}_x$. 
Although the quantumness $\hbar$ is only heuristic, being a Schrödinger equation makes the advection equation amenable to exponential speedup on quantum computers. 
The difficulty is that eigen functions of $\hat{p}_x$ are Fourier modes, so the unitary dynamics generated by $\hat{p}_x$ are oscillatory. The oscillatory behavior is especially pronounced when $A$ has steep gradients. The oscillatory behavior is less preferable than dissipative behavior in applications where capturing shocks is important. However, dissipation is not unitary. The goal of this paper is to develop a quantum algorithm for enacting the advection operator $d_t = \partial_t + c\partial_x$ that avoids spurious oscillations while retaining an exponential quantum speedup.

\subsection{Upwind Discretization}
In classical computing, upwind discretization is a well-known method for suppressing spurious oscillations \cite{Courant52,leveque_finite_nodate}. We adopt the first-order upwind  scheme and turn it into a quantum algorithm. For the partial derivative in time, forward Euler scheme approximates
$\partial_t A\approx \frac{1}{\Delta t}[A(x,t+\Delta t)-A(x,t)]$.
The partial derivative in space for $c > 0$ is approximated by $\partial_x A \approx \frac{1}{\Delta x}[A(x,t)-A(x-\Delta x,t)]$, 
whereas for $c < 0$, the partial derivative is instead approximated by $\partial_x A \approx \frac{1}{\Delta x}[A(x+\Delta x,t)-A(x,t)]$.
This first-order upwind scheme respects causality by only taking information from the upwind side of the advection. The algorithm is stable, including near shock-like discontinuities, when the Courant–Friedrichs–Lewy (CFL) condition $|c|\Delta t \leq \Delta x$ is satisfied \cite{courant_partial_1956}. 
Denoting the discretized waveform $a_i^n\approx A(x=x_0 + i\Delta x, t=t_0 + n\Delta t)$, the first-order upwind scheme for $c>0$ is 
\begin{equation}
    \frac{a_i^{n+1} - a_i^n}{\Delta t} + c\frac{a_i^{n} - a_{i-1}^n}{\Delta x} = 0,
\end{equation}
where $i$ is the spatial index and $n$ is the temporal index. 
Up to an overall normalization factor that can be predetermined, the waveform at time $n$ can be encoded as a quantum state vector $\vec{a}^n$. If we bound the finite spatial state into a box of length $L$ and assume periodic boundary conditions, the discretized equation can be rearranged as the following matrix equation
\begin{equation}
\vec{a}^{n+1} = 
    \begin{bmatrix}
    1-f & 0 & \dots & 0 & f \\
    f & 1-f & 0 &  & 0 \\
    0 & f & 1-f & \ddots & \vdots \\
    \vdots & \ddots & \ddots & \ddots & 0 \\
    0 & \dots & 0 & f & 1-f \\
    \end{bmatrix} \vec{a}^n,
\label{positive_recurrence}
\end{equation}
where $f = |c|\frac{\Delta t}{\Delta x}$ is the CFL number, which must satisfy $0 \leq f \leq 1$ for stability. When $c$ and $\Delta x$ are given, the CFL condition imposes the maximum time step size $\Delta t$ at which the discretized waveform can be advected.
In the case $c<0$, the matrix equation is instead 
\begin{equation}
\vec{a}^{n+1} = 
    \begin{bmatrix}
    1-f & f & 0 & \dots & 0 \\
    0 & 1-f & f & \ddots & \vdots \\
    \vdots & 0 & 1-f & \ddots & 0 \\
    0 &  & \ddots & \ddots & f \\
    f & 0 & \dots & 0 & 1-f \\
    \end{bmatrix} \vec{a}^n.
\label{negative_recurrence}
\end{equation}
We denote these recurrence matrices as $\hat{T}^+$ and $\hat{T}^-$, for positive and negative $c$, respectively. Neither matrix is unitary unless $f=0$ or $f=1$.
The recurrence relations in Eqs.~(\ref{positive_recurrence}) and (\ref{negative_recurrence}) describe how the state vector evolves in time. By applying $\hat{T}^+$ or $\hat{T}^-$ once, the algorithm enacts the advection operator for one time step.

\subsection{Quantum Implementation of $\hat{T}^+$ and $\hat{T}^-$}

The matrices $\hat{T}^+$ and $\hat{T}^-$ are not unitary, so they cannot be directly implemented on a quantum computer. However, they can be written as a linear combination of unitary operators given by 
\begin{equation}
\hat{T}^+ = (1-f)
    \begin{bmatrix}
    1 & 0 & 0 & \dots & 0 \\
    0 & 1 & 0 & \ddots & \vdots \\
    \vdots & 0 & 1 & \ddots & 0 \\
    0 &  & \ddots & \ddots & 0 \\
    0 & 0 & \dots & 0 & 1 \\
    \end{bmatrix} + f
    \begin{bmatrix}
    0 & & \dots & 0 & 1 \\
    1 & 0 & & & 0 \\
    0 & 1 & 0 & & \vdots \\
    \vdots &  & \ddots & \ddots & 0 \\
    0 & \dots & 0 & 1 & 0 \\
    \end{bmatrix} = 
    (1-f)\hat{I} + f\hat{U}^+,
    \label{T+_linear_decomp}
\end{equation}
\begin{equation}
\hat{T}^- = (1-f)
    \begin{bmatrix}
    1 & 0 & 0 & \dots & 0 \\
    0 & 1 & 0 & \ddots & \vdots \\
    \vdots & 0 & 1 & \ddots & 0 \\
    0 &  & \ddots & \ddots & 0 \\
    0 & 0 & \dots & 0 & 1 \\
    \end{bmatrix} + f
    \begin{bmatrix}
    0 & 1 & 0 & \dots & 0 \\
    0 & 0 & 1 & \ddots & \vdots \\
    \vdots & 0 & 0 & \ddots & 0 \\
    0 &  & \ddots & \ddots & 1 \\
    1 & 0 & \dots & 0 & 0 \\
    \end{bmatrix} = 
    (1-f)\hat{I} + f\hat{U}^-,
    \label{T-_linear_decomp}
\end{equation}
where $\hat{I}$ is the identity matrix. 
In these equations, we introduce the shifting operators $\hat{U}^\pm$, which shift every point in the state vector by distance $\pm\Delta x$. In other words, $\hat{U}^+$ ($\hat{U}^-$) shifts the state to the right (left) by one grid point. The two shifting operators are transpose of each other and are both unitary.

\subsubsection{Linear Combination of Unitaries}
Linear combinations of unitaries (LCUs) \cite{childs_hamiltonian_2012} can be implemented on quantum computers using additional (ancilla) qubits. The procedure involves four steps: prepare the ancilla qubits to the square roots of the LCU weights, selectively apply each unitary to data qubits by controlling on the ancilla, unprepare the ancilla qubits, and measure the ancilla qubits. If ancilla qubits return to the $\ket{0}$ state, the data qubits then collapse to the desired state.
In our case, $\hat{T}^\pm$ is a linear combination of just two unitaries: the identity matrix, which requires no operation, and a shifting matrix $\hat{U}^{\pm}$. Following the LCUs protocol, the quantum circuit that implements $\hat{T}^\pm$ is shown below, where the data qubits are collectively labeled as $\ket{\Psi}$. For $N$ spatial grid points, the number of data qubits needed is $n=\log_2 N$. 
\begin{equation}
\nonumber
\begin{quantikz}
  \lstick[3]{\ket{\Psi}}&&&\gate[3]{\hat{U}^\pm}&&& \\
  &&&&&& \\
  &&&&&& \\
  \lstick{ancilla}&& \gate[1]{\hat{P}} & \ctrl{-1} & \gate[1]{\hat{P}^{\dagger}}&\meter{}&\\
\end{quantikz}
\label{T_circuit_1}
\end{equation}
First, a single ancilla qubit is prepared to the state $\sqrt{1-f}|0\rangle + \sqrt{f}|1\rangle$ using the $\hat{P}$ operator. The $\hat{P}$ operator is a single-qubit rotation, which can be easily implemented on quantum hardware. 
Second, the select operator $\hat{S}$ applies $\hat{U}^\pm$ if the ancilla is in $|1\rangle$ state. The select operator is a controlled multi-qubit operation, whose gate-level decomposition is discussed below. 
Third, the inverse $\hat{P}$ operator is applied to the ancilla, which rotates the ancilla back to its original state. 
Finally, the ancilla is measured, which introduces non-unitarity into the otherwise unitary quantum circuit. Depending on the outcome of the ancilla measurement, $\ket{\Psi}$ collapses into a success or a failure state.

Mathematically, the first three steps of LCUs before the final measurement are unitary. When acting on the tensor product state $|0\rangle|\Psi\rangle$, the outcome of the overall unitary operation is  
\begin{align}
    \hat{P}^{\dagger}\hat{S}\hat{P}|0\rangle|\Psi\rangle &= |0\rangle[(1-f)\hat{I} + f\hat{U}^\pm]|\Psi\rangle \notag \\
    &+ |1\rangle\sqrt{f(1-f)}(\hat{U}^\pm - \hat{I})|\Psi\rangle.
\label{LCU_math}
\end{align}
At this point, the ancilla and data qubits are entangled. Measuring the ancilla disentangles them and collapses the data qubits into two possible quantum states
\begin{equation}
\begin{cases}
    \alpha^\pm[(1-f)\hat{I} + f\hat{U}^\pm]\ket{\Psi}, & \text{ancilla} = \ket{0} \\
    \beta^\pm(\hat{U}^\pm - \hat{I})|\Psi\rangle, & \text{ancilla} = \ket{1} \\
\end{cases}
\label{LCU_result}
\end{equation}
where $\alpha^\pm$ and $\beta^\pm$ are constants such that the final state of the data qubits is normalized. Comparing the possible outcomes to $\hat{T}^\pm$ [Eqs.~(\ref{T+_linear_decomp}) and (\ref{T-_linear_decomp})], the desired operation is achieved only if the measurement finds the ancilla in the $\ket{0}$ state. 
In other words, the ancilla serves as a flag, and if the flag is $\ket{0}$, the data qubits are in a success state. On the other hand, if the flag is $\ket{1}$, the data qubits are in a failure state. 
Because quantum measurement is probabilistic, the implementation of $\hat{T}^\pm$ using LCUs introduces a failure probability at each time step.

\subsubsection{Gate Decomposition \label{sec:U_decomposition}}
In LCUs, the select operator $\hat{S}$ is an oracle. Here we show that this ``black box" operation can be decomposed as multi-controlled NOT gates. First, consider the gate decomposition of the shifting operators $\hat{U}^\pm$. 
For a single data qubit, $\hat{U}^\pm$ are simply the NOT gate, namely, the Pauli $X$ gate, which flips the qubit $\ket{0}\leftrightarrow\ket{1}$. 
For two data qubits, $\hat{U}^\pm$ are Controlled NOT (CNOT) gates. 
In the case of four data qubits, $\hat{U}^\pm$ are decomposed by 
\begin{equation}
\nonumber
\hat{U}^+ = 
\begin{quantikz}
  &\ctrl{1}&\ctrl{1}&\ctrl{1}&\targ{}& \\
  &\ctrl{1}&\ctrl{1}&\targ{}&& \\
  &\ctrl{1}&\targ{}&&& \\
  &\targ{}&&&&\\
\end{quantikz}
\label{U+_decomp}
\end{equation}
\begin{equation}
\nonumber
\hat{U}^- = 
\begin{quantikz}
  &\targ{}&\ctrl{1}&\ctrl{1}&\ctrl{1}&\\
  &&\targ{}&\ctrl{1}&\ctrl{1}& \\
  &&&\targ{}&\ctrl{1}& \\
  &&&&\targ{}& \\
\end{quantikz}
\label{U-_decomp}
\end{equation}
In the circuit diagrams above, the $\oplus$ symbol denotes the target qubit, whose state is flipped when the control qubits are all in $\ket{1}$ state, which are denoted by solid black dots. 
In Appendix A, we prove by induction that for $n$ qubits, the decomposition pattern continues: a cascade of $n$ multi-controlled NOT gates of decreasing ($\hat{U}^+$) or increasing ($\hat{U}^-$) depth.

In the select operator $\hat{S}$, the ancilla is just an additional control qubit. 
The quantum circuit below implements one full advection time step for three data qubits (eight spatial grid points) with a positive advection speed. 
\begin{equation}
\nonumber
\begin{quantikz}
  \lstick[3]{\ket{\Psi}}&&\ctrl{1}&\ctrl{1}&\targ{}&&& \\
  &&\ctrl{1}&\targ{}&&&& \\
  &&\targ{}&&&&& \\
  \lstick{ancilla}& \gate[1]{\hat{P}} & \ctrl{-1} & \ctrl{-2} & \ctrl{-3} & \gate[1]{\hat{P}^{\dagger}} & \meter{}&\\
\end{quantikz}
\end{equation}
Each multi-controlled gate involving $n$ qubits can be decomposed further into $O(n)$ two-qubit gates \cite{barenco_elementary_1995}. 
Moreover, on quantum computing platforms like ion traps, multi-controlled gates may be implemented natively without further gate decomposition \cite{grzesiak_efficient_2020}. 
The simple structure of $\hat{T}^\pm$ makes it realizable on near-term quantum hardware.
For $N=2^n$ spatial grid points, the quantum advection algorithm requires $O(n)$ memory and $O(\text{poly}\, n)$ operations per time step, which is exponentially more efficient than classical advection algorithms which require $O(N)$ memory and $O(N)$ operations per time step.

\subsection{Demonstration Using State Vector Simulator\label{sec:demonstration_advection}}
\begin{figure}[H]
	\centering
	\includegraphics[width=0.8\linewidth]{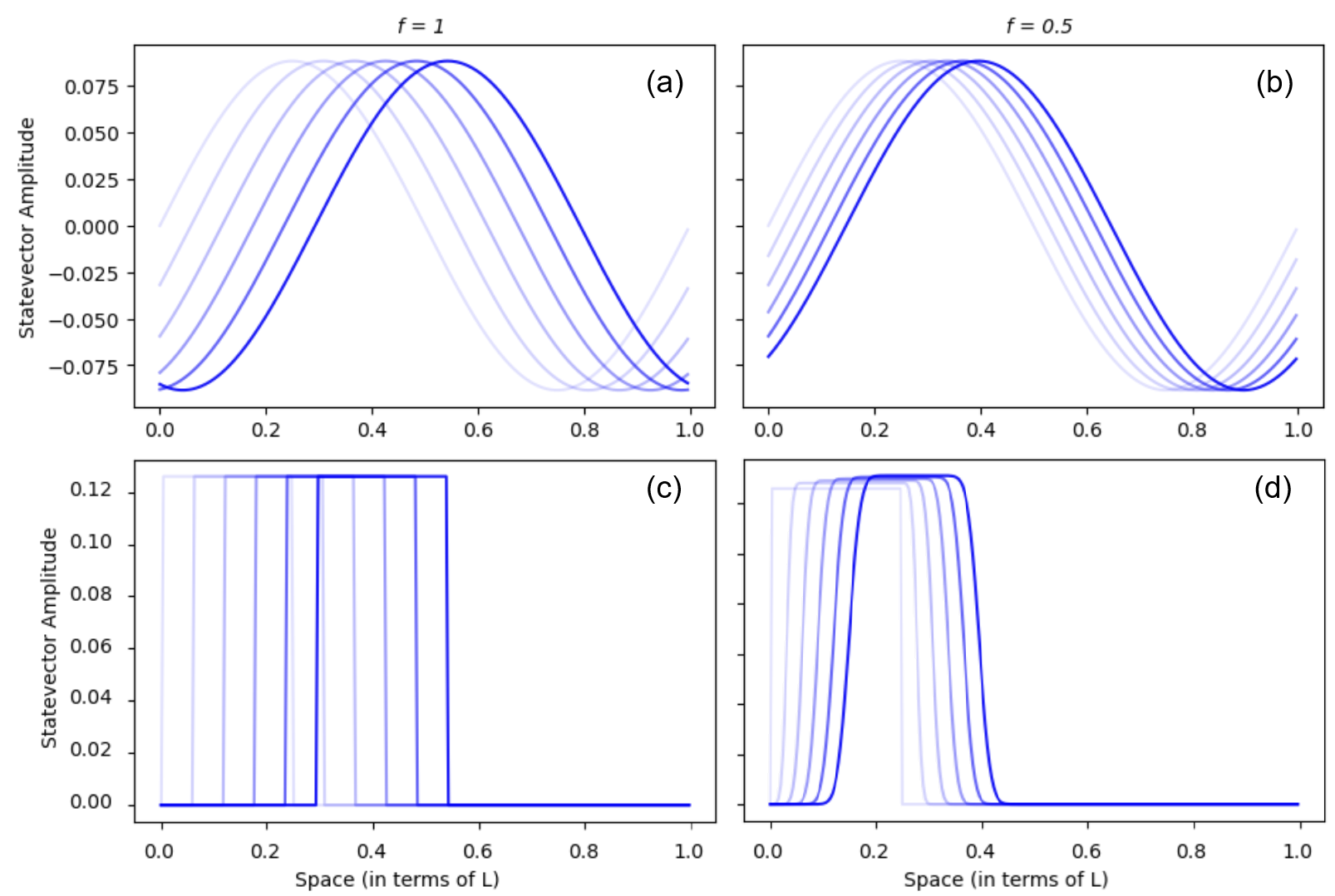}
	\caption[Advection Demonstration]{
		Using Qiskit's state vector simulator, the upwind quantum advection operator is repeatedly applied to a sine (a)-(b) and a square (c)-(d) waveform on $8$ data qubits with a positive speed. In each figure, lines are separated by $15$ time steps, and later time steps are represented by darker lines. When $f=1$ (left), the algorithm is exact. When $0<f<1$, the algorithm introduces numerical diffusion, whose worst case ($f=1/2$) is shown on the right. The advection algorithm preserves the norm of waveforms and their spatial integrals. 
	}
	\label{Advection_h}
\end{figure}

We demonstrate our shock-capturing quantum advection algorithm using Qiskit's state vector simulator \cite{qiskit2024}. The simulator is a classical linear algebraic calculator without shot noise or hardware infidelities. After each time step, we project the quantum state to the success state, which is then continued to the next time step.
Figure \ref{Advection_h} shows results after performing advection for multiple time steps when the initial waveform is a sine (upper) and a square (lower). 
The advection algorithm propagates the waveforms to the right with the expected constant speed. 
When the CFL number $f=1$, the advection moves the waveform by exactly one grid point per time step, so the numerical advection exactly matches the analytical result. 
When $f=1/2$, the advection moves the waveform by half a grid point per time step.
Because the discrete grid has the most difficulty in capturing half a grid point of movement, the upwind algorithm introduces the worst-case numerical diffusion. For the sine waveform, the numerical diffusion is exactly canceled by the renormalization every time step. For the square waveform, no such cancellation occurs, and the dissipation is clearly observed. Unlike previous unitary quantum algorithms that introduce large spurious oscillations for the square waveform, our LCUs algorithm causes the square to round off. The height of the rounded square increases because the waveform remains normalized. 
Our LCUs algorithm is applicable to any initial waveform as long as the CFL condition is satisfied.

\section{Failure Recovery Algorithm}
The dissipative upwind scheme is non-unitary, and the LCUs implementation fails when the ancilla is measured in the $\ket{1}$ state. Suppose a lower bound of the success probability per step is $p$, then after compounding $S$ time steps, a naive lower bound of the overall success probability is $p^S$, which is exponentially diminishing in $S$. The implication is that in order to successfully advect for $S$ consecutive steps, an upper bound for the expected number of trials needed is $1/p^S$, which would imply an exponential complexity in time. In Sec.~\ref{sec:efficiency}, we show that the time complexity is actually far better. Nevertheless, if we can recover from a failure state rather than restarting from the beginning, the overall success probability is improved.  

\subsection{Failure State}
When the ancilla is measured in the $\ket{1}$ state, according to Eq.~(\ref{LCU_result}), the failure state is $\beta^\pm (\hat{U}^\pm - \hat{I})\ket{\Psi}$, where $\beta^\pm$ is a normalization factor. Using the explicit forms of the shifting operators, the failure state for $c>0$ is
\begin{equation}
    \beta^+(\ket{\Psi(x+\Delta x)} - \ket{\Psi(x)}), 
\end{equation}
and for $c < 0$, the failure state is 
\begin{equation}
    \beta^-(\ket{\Psi(x-\Delta x)} - \ket{\Psi(x)}).
\end{equation}
These failure states are proportional to $\pm\Delta x \partial_x \Psi$, where the numerical derivative of the state vector is approximated using either forward or backward finite difference. 
For a bounded $|\partial_x\Psi|$, the norm of the failure state, and therefore the failure probability, is proportional to $\Delta x^2$, which goes to zero when the spatial resolution refines. 
Moreover, because the failure state is the derivative of the original state, when the LCUs algorithm fails, the original state can be recovered from the failure state by performing an indefinite integration.

\begin{figure}[!b]
    \begin{center}
    \includegraphics[width=100mm]{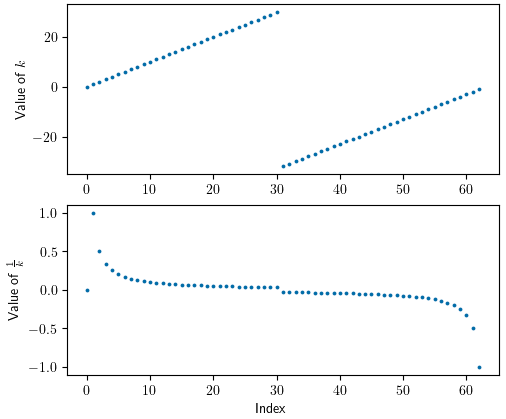}
    \end{center}
    \caption[Behavior of k]{
        Failure recovery is performed in the momentum space after quantum Fourier transform, whose natural ordering of $k$ is unshifted (upper). The failure recovery scheme requires $1/k$, which goes to zero towards the middle of the box (lower). The index starts from zero and the example uses $N=64$ grid points.
	}
   \label{k_shape}
\end{figure}

\subsection{Quantum Indefinite Integration\label{sec:quantun_integral}}
To perform indefinite integration of a given but unknown quantum state, we develop a scheme based on quantum Fourier transform (QFT) \cite{coppersmith_approximate_2002}, which is known to offer exponential speedup over classical fast Fourier transform (FFT). 
Denote the failure state $\Phi(x) = \beta\frac{d}{dx}\Psi(x)$, where the unknown factor $\beta$ is such that both $\Phi(x)$ and $\Psi(x)$ are normalized. 
Taking the Fourier transform $\int_{-\infty}^{\infty} dx\, e^{-ikx}$ on both sides, 
\begin{equation}
    \label{eq:QFT}
    \text{QFT}[\Phi] (k) = -ik\beta\; \text{QFT}[{\Psi}](k).
\end{equation}
Here, $k$ is written as a continuous variable for convenience. The discretized version is analogous, but care must be taken to map the integer index to the discrete $k$ values, as shown in Fig.~\ref{k_shape}.
For continuous Fourier transform, $k$ increases linearly from $-\infty$ to $+\infty$. However, QFT is typically defined on spatial grid points $[0,N)$ in a finite box with implied periodic extensions. So, the natural ordering of the $k$ grid first increases linearly from $0$ to $N/2$, flips its sign, and then continues to increase from $-N/2$ to zero. 
Rearranging Eq.~(\ref{eq:QFT}), given the failure state $\Phi$, the original state $\Psi$ can be recovered by
\begin{equation}
    \Psi(x) = \text{QFT}^{-1}\left[\frac{\text{QFT}[\Phi]}{-ik\beta }\right],
\end{equation}
which effectively achieves the indefinite integral $\Psi(x) = \frac{1}{\beta}\int dx\,\Phi(x)$.
In other words, the indefinite integral of $\Phi(x)$ can be computed by applying QFT, multiplying by $\frac{1}{-ik\beta}$, and then applying inverse QFT.

In order to avoid division by $k=0$ in our indefinite integration scheme, we impose the constraint that the spatial integral of the initial waveform must be zero, namely, $\int dx\, \Psi(x)=0$ in the computational box. Under this constraint, the $k=0$ Fourier mode has zero amplitude and can thus be ignored. 
This constraint can always be satisfied by shifting $\Psi$ by a constant offset. 
The offset does not change in time because first-order upwind advection conserves the spatial integral of $\Psi$. Therefore, before the quantum simulation starts, the constant can be precomputed and stored classically, and after the quantum simulation finishes, the offset can be restored if needed during classical post processing. 
Another way to understand this constraint is that indefinite integration is only unique up to a constant. Information about the constant is lost in the derivative state and cannot be recovered. Nevertheless, as quantum states, both $\Psi(x)$ and $\Phi(x)$ must always remain square normalized, which means that when shifted by a wrong constant, the recovered $\Psi(x)$ is distorted from its original waveform.
The constraint $\int dx\, \Psi(x)=0$ is a protocol for setting the constant offset, such that the waveform $\Psi(x) = \int dx\, \partial_x \Psi(x)$ remains unchanged after passing it through the failure derivative and then the recovery integral.

\subsubsection{Multiplication by $i/k$}
The quantum circuits for QFT and its inverse are well known \cite{coppersmith_approximate_2002}. 
The only unsolved part of our indefinite integration algorithm is multiplying the state vector by $\frac{1}{-ik\beta }$. 
In grid units and excluding the point $k=0$, because $|\frac{1}{k}| \leq1$, multiplication by $\frac{1}{k}$ does not preserve the norm of the state vector, so the operation is not unitary and must be performed probabilistically on quantum computers.
One method to implement non-unitary operators on quantum computers is to create a block encoding \cite{low_hamiltonian_2019}. Adding an additional ancilla qubit, any non-unitary operator $\hat{A}$ can be encoded into a larger operator of the form
\begin{equation}
    \begin{bmatrix}
        \hat{A} & * \\
        * & *
    \end{bmatrix},
\end{equation}
where *'s indicate matrix blocks of the same dimension as $\hat{A}$. These matrices are chosen such that the entire operator is unitary. After the enlarged operator is applied, a measurement is made on the ancilla qubit. If the measurement finds the ancilla in the $\ket{0}$ state, the state of the data qubits collapses to the desired state resulting from the application of $\hat{A}$.

For multiplication by $\frac{1}{-ik \beta }$, the desired matrix is $\hat{A}=\frac{1}{-ik \beta }\hat{I}$, where $\hat{I}$ is the identity matrix.
One possible choice is to set both diagonal quadrants to $\frac{1}{-ik\beta }\hat{I}$. The off-diagonal quadrants can then be solved by requiring the matrix to be unitary, which gives
\begin{equation}
    \begin{bmatrix}
        \frac{1}{-ik\beta }\hat{I} & \sqrt{1-(\frac{1}{\beta k})^2}\hat{I} \\
        \sqrt{1-(\frac{1}{\beta k})^2}\hat{I} & \frac{1}{-ik\beta }\hat{I}
    \end{bmatrix}.
\end{equation}
The unknown normalization constant $\beta$ is in fact unnecessary, because after the application of $\hat{K}$ and the measurement of the ancilla qubit, the data qubits are automatically normalized. So, $\beta$ does not affect the final result, as long as unitarity is preserved. A matrix can only be unitary if each of its elements has a magnitude less than or equals to $1$. Because the maximum value of $1/k$ is $1$, $\beta$ must be greater than or equal to $1$. Additionally, the probability of a successful measurement of the ancilla is proportional to the norm of the desired state $|\frac{1}{-ik \beta }\ket{\Psi}|^2$. Therefore, $\beta$ should be as small as possible to maximize this success probability. Thus, we set $\beta=1$ and the operator that multiplies the data qubits by $i/k$ is given in matrix form by
\begin{equation}
    \hat{K}=
    \begin{bmatrix}
        \frac{i}{k}\hat{I} & \sqrt{1-\frac{1}{k^2}}\hat{I} \\
        \sqrt{1-\frac{1}{k^2}}\hat{I} & \frac{i}{k}\hat{I}
    \end{bmatrix}.
\end{equation}
The magnitudes of diagonal elements can be read out from Fig.~\ref{k_shape} (lower). Notice that the first diagonal element of the matrix is zero, but the last diagonal element is nonzero, which is a consequence of the natural ordering of $k$.

\subsubsection{Gate Decomposition of $\hat{K}$ \label{sec:decompose_K}}
To implement the unitary matrix $\hat{K}$ on quantum computers, we decompose it into elementary gates. Because $\hat{K}$ has entries other than $0$ or $1$, its gate decomposition is more complicated than $\hat{U}^\pm$. While $\hat{U}^\pm$ is decomposed into multi-controlled $X$ gates, we decompose $\hat{K}$ into multi-controlled $R_X$ gates, which are defined as 
\begin{equation}
    \hat{R}_X(\theta)=
    \begin{bmatrix}
        \cos(\theta/2) & -i\sin(\theta/2) \\
        -i\sin(\theta/2) & \cos(\theta/2)
    \end{bmatrix},
\end{equation}
where $\theta$ is an arbitrary rotation angle \cite{barenco_elementary_1995}.
It is worth clarifying that $X$ has nothing to do with the spatial dimension $x$ for the advection equation. Here, $\hat{R}_X(\theta)$ is a rotation around the $X$ axis of the Bloch sphere, which lives in the Hilbert space of a single qubit, rather than the configuration space of the advection.

Building upon single-qubit $\hat{R}_X(\theta)$ rotations, multi-controlled rotation gates can be defined, similar to a multi-controlled NOT (X) gate. 
To simplify our notations, we suppress $X$, with the implied understanding that all rotations are around the $X$ axis. 
We denote multi-controlled rotation gates by $\hat{R}_{t, q=b}^{\theta}$, where $t$ is the target qubit, $q$ is the set of control qubits, $b$ is a bit string of the same length as $q$, and $\theta$ is the desired rotation angle. 
The $\hat{R}_{t, q=b}^{\theta}$ gate rotates the target qubit $t$ by angle $\theta$ if and only if the control qubits $q$ are equal to the bit string $b$.
For example, consider a quantum circuit with four qubits indexed from 0 to 3. The $\hat{R}_{2, [0,1,3]=[1,1,0]}^{\frac{\pi}{2}}$ gate rotates qubit 2 by $\frac{\pi}{2}$ only if qubits 0, 1, and 3 are in $\ket{1}, \ket{1}$, and $\ket{0}$ states, respectively. Otherwise, the state of qubit 2 is unchanged.

\begin{figure}[H]
  \centering
  \includegraphics[width=0.9\linewidth]{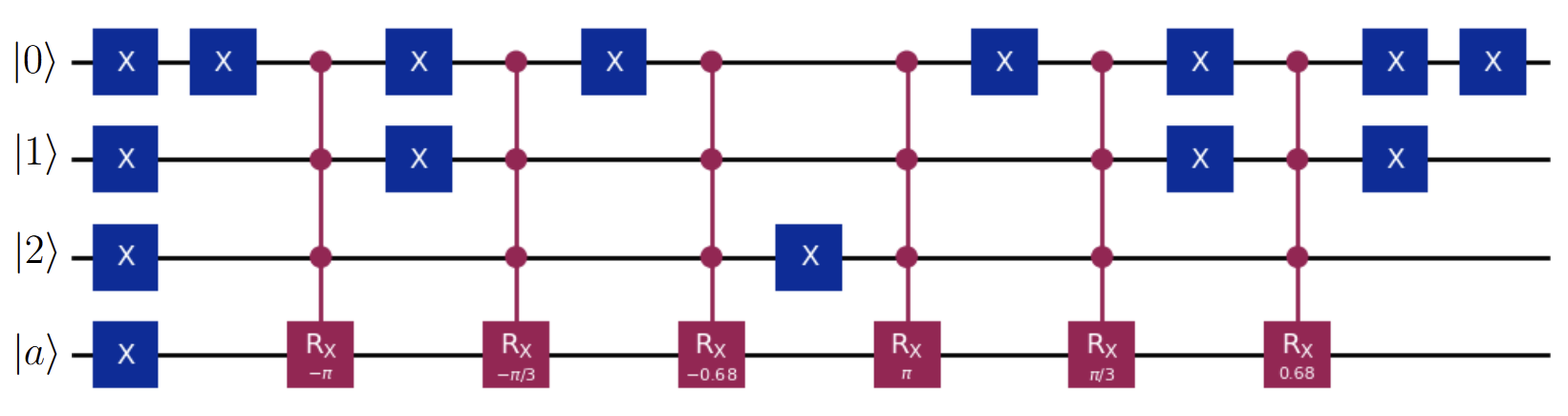}
  \caption[$\hat{K}$ Quantum Circuit]{
	Quantum circuit for the failure recovery operator $\hat{K}$ in the case of three data qubits and one ancilla qubit $\ket{a}$. $X$ (NOT) gates are initially applied to all qubits, setting the bit string to all 0's. The $X$ gate on the ancilla creates a template that resembles $\hat{K}$. We then count up in binary using $X$ gates, applying appropriate multi-controlled $R_X$ rotations in between. At the halfway point, the single X gate on the last data qubit switches to the end of the $\hat{K}$ matrix, from which the matrix is edited by the remaining gates in reverse order. 
	The final X gates reset the bit string to its original values.
	}
  \label{K_circuit}
\end{figure}

In the case of $\hat{K}$, we have a quantum circuit with $n$ data qubits and $1$ ancilla qubit $a$. 
Following our notations, the operator $\hat{R}_{a, [0:n-1]=[10...0]}^{\theta}$ applies a rotation on the ancilla if the 0-th data qubit is in the excited state while all other data qubits are in the ground state. In matrix form, 
\begin{equation}
   \hat{R}_{a, [0:n-1]=[10...0]}^{\theta}=
\begin{bmatrix}
   1 & 0 & \dots & 0 & 0 & 0 & \dots& 0 \\
   0 & -i\sin(\theta/2) & &  \vdots & 0 & \cos(\theta/2) &&\vdots \\
   \vdots & & \ddots & 0 & \vdots &&\ddots&0 \\
   0 & \dots & 0 & 1 & 0 &\dots& 0 & 0 \\
   0 & 0 &\dots& 0 & 1 & 0 & \dots & 0 \\
   0 & \cos(\theta/2) && \vdots & 0 & -i\sin(\theta/2) & & \vdots \\
   \vdots && \ddots & 0 & & & \ddots & 0 \\
   0 &\dots& 0 & 0 & 0 & \dots & 0 & 1 \\
\end{bmatrix}.
\label{second_edit}
\end{equation}
This operator edits the second diagonal entry of each quadrant. Following the same pattern, the operator $\hat{R}_{a, [0:n-1]=b}^{\theta}$ edits the $(b+1)$-th entry of each quadrant. 
Looping over $b$ and applying multi-controlled rotations with appropriate angles, we can construct $\hat{K}$ one diagonal entry at a time. 
Instead of editing from an initial identity operator, applying an initial $X$ gate to the ancilla starts the operator from a template operator $X\otimes\hat{I}$ that closely resembles $\hat{K}$.
The template is the asymptotic form of $\hat{K}$ when $k\rightarrow\infty$, where the main diagonal goes to $0$ and the two off-diagonal blocks go to $1$.
The template also ensures that the first diagonal entry of the diagonal blocks are $0$ as desired.
Figure~\ref{K_circuit} shows the full quantum circuit that implements $\hat{K}$ in the example of three data qubits.

\subsection{Final Half-Step Advection}
The failure state is an upwind derivative of the un-advected waveform. The derivative effectively shifts the waveform by $\Delta x/2$ in the $-c$ direction. Thus, after indefinite integration, the resulting waveform is $\Delta x/2$ behind the original waveform. To fully recover the state, we perform a final advection step with $f=1/2$. 
The failure recovery succeeds if both the indefinite integration and the final half-step advection succeed. From the recovered state, the advection algorithm can resume and continue to the next step, instead of discarding the failure state and restarting from the beginning of the simulation.

\section{Algorithm Efficiency\label{sec:efficiency}}
Classically, numerical advection schemes can scale no better than $O(N)$ per time step, where $N=2^n$ is the number of grid points. This limit in efficiency comes from the fact that each data point must be moved at each time step for advection to occur. Therefore, if our quantum algorithm scales better than $O(N)$ per time step, quantum speedup is achieved.

\subsection{Advection Efficiency\label{sec:advection_efficiency}}
In Sec.~\ref{sec:U_decomposition}, we show that one advection step for $n$ data qubits is implemented by two single-qubit rotations on the ancilla, $n$ multi-controlled NOT gates, and a single-qubit measurement. 
Single-qubit operations on the ancilla are straightforward, so most of the complexity comes from the multi-controlled NOT gates.
On a generic quantum computing platform, multi-controlled gates cannot be implemented natively without further decomposition. Barenco et al. describe a decomposition of multi-controlled NOT gates that uses $O(n)$ elementary gates \cite{barenco_elementary_1995}. Since there are $n$ such multi-controlled gates, the number of two-qubit gates needed for each advection step is $O(n^2)=O(\log_2(N)^2)$.
Assuming the LCUs advection is successful, the quantum algorithm is exponentially faster than classical algorithms. 
To advect a waveform across a desired distance, the total cost is the cost per time step multiplied by the desired number of steps. 
Due to the CFL condition, both classical and quantum advection can move no more than one grid point per time step. Thus, to advect a distance comparable to the simulation box size, $O(N)$ time steps must be taken.
For classical algorithms, the total number of operations needed is therefore $O(N^2)$, while for our quantum algorithm, the total complexity is $O(N\log(N)^2)$. The origin of the significant quantum speedup is the exponential decrease in the cost per time step. In the following, we focus on the complexity per step.

At each time step, the LCUs advection can fail when the ancilla is measured in the $\ket{1}$ state. 
Without failure recovery, we need to restart the simulation, running multiple trials until the algorithm succeeds consecutively to reach the final desired advection distance. 
Repeating trials increases the complexity of the algorithm. 
To incorporate effects of failures, we compute the expected number of operations needed to advect some desired distance $d/L$ and then divide by the number of time steps required to reach that distance. This metric represents the average number of operations per time step for any desired advection distance and is physically meaningful. 
Below, we prove that a non-zero failure probability does not affect the asymptotic complexity, maintaining the exponential speedup in the large $n$ limit.

To estimate the complexity, we define an attempt as a string of consecutive time steps. An attempt can either end in success, when the algorithm reaches its distance goal $d/L$, or failure, when a failed time step is encountered. If failure occurs, the algorithm restarts from the beginning. We define $P_{\text{success}}$ as the probability of a successful attempt. Because each step has an independent success probability, 
\begin{equation}
    P_{\text{success}} = \prod_{t=1}^{S}(1-p_t),
    \label{general_P_success}
\end{equation}
where $p_t$ is the failure probability at time step $t$ and $S$ is the total number of time steps needed to reach the goal. 
From the CFL condition,
\begin{equation}
    \label{eq:advection_time}
    S = \frac{d}{f\Delta x} = 2^n\frac{d}{Lf},
\end{equation}
which is rounded up to the nearest integer. 
Because $P_{\text{success}}$ is identical and independent across attempts, the expected number of attempts to succeed at least once is $1/P_{\text{success}}$. 
Each attempt costs $O(2^n)$ operations to prepare the initial state \cite{mottonen_transformation_2005}, and then costs $O(n^2)$ operations per step. 
We set the number of steps per attempt to $S$, which is an upper bound because an attempt could have been terminated by a failure prior to $S$. 
Putting everything together, the expected number of operations for an advection distance $d/L$ is $O\left(\frac{2^n}{P_{\text{success}}}(1 + \frac{d}{Lf} n^2)\right)$.
Dividing by $S$, the expected number of operations per time step is
\begin{align}
    O\left(\frac{1}{P_{\text{success}}}(\frac{Lf}{d} + n^2)\right).
    \label{exact_complexity}
\end{align}
In the asymptotic limit of large $n$, the constant $\frac{Lf}{d}$, which corresponds to the cost of a reset, is negligible. 
The reset cost is only significant at low qubit numbers and short distances. Dropping the reset cost, we get a simplified complexity of $O(n^2/P_{\text{success}})$. 
The remaining task is to analyze how $P_{\text{success}}$ depends on $n$, $d/L$, and other parameters.

\begin{figure}[H]
    \begin{center}
	\includegraphics[width=0.9\textwidth]{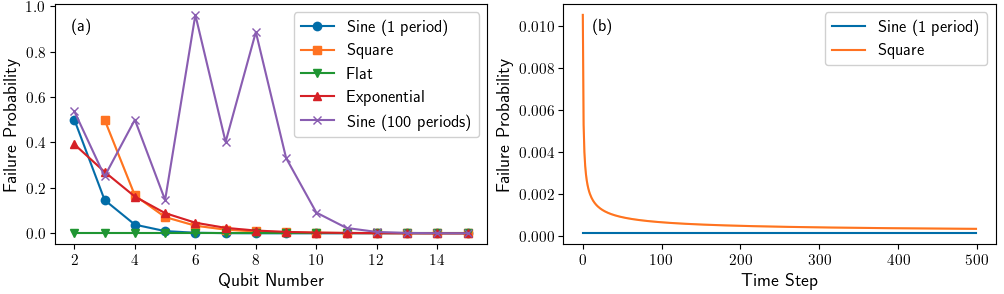}
    \end{center}
    \caption[Advection Failure Probabilities]{
	(a) The failure probability decreases exponentially with increasing resolution for all waveforms after transient behaviors at small numbers of qubits. 
    (b) At a given resolution (8 qubits), the failure probability decreases with time due to numerical diffusion that smooths the waveforms as the simulations progress.    
    The failure probability is calculated classically according to Eq.~(\ref{failure_state}) with the worst-case $f=1/2$.
    }
\label{failure_probs}
\end{figure}

The failure probability $p_t$ at time step $t$, according to Eq.~(\ref{LCU_math}), is given by the norm of the failure state 
\begin{equation}
    p_t = |\sqrt{f-f^2}(\hat{U}^{\pm} - \hat{I})\ket{\Psi_t}|^2.
\label{failure_state}
\end{equation} 
Because $\hat{U^\pm}$ shifts the state by $\pm\Delta x$, the failure probability is proportional $|\Delta x\partial_x\Psi|^2$ multiplied by $f-f^2$, which is maximized when $f=1/2$. Figure \ref{failure_probs} shows failure probabilities calculated classically using Eq.~(\ref{failure_state}) for a variety of initial waveforms.
For each waveform, the failure probability decreases with the number of qubits after some transient behavior, which is function dependent. An extreme example of this transient behavior is the 100-period sine wave, which is unresolved at lower qubit numbers.
After a waveform is adequately resolved, the transient behavior ceases and the failure probability decreases exponentially.
For functions with bounded $|\partial_x\Psi|^2$, the failure probability per step
$\Delta x^2\propto 2^{-2n}$ decreases faster than the increase of $S\sim 2^n$, so effects of failures diminish exponentially in the large $n$ limit. Other types of functions have different failure probability scalings, as discussed later in this section.   
Additionally, Fig.~\ref{failure_probs}(b) shows that the failure probability decreases with time. The rapid decrease is due to numerical diffusion inherent to the upwind discretization scheme \cite{leveque_finite_nodate}. Diffusion smooths the waveform, reducing the norm of the derivative state $|\partial_x\Psi|^2$, and thus further reducing the probability of failure as simulations progress.

Combining numerical diffusion with the exponential decrease of failure probability with $n$, we now show that for all $L^2$ functions, $P_{\text{success}}$ converges to $1$ when $n \to \infty$. 
Any initial square integrable function $\Psi \in L^2$ can be expanded in terms of orthonormal Fourier modes $\phi_m$.
For normalized $\Psi$ discretized on $N$ grid points, the Fourier expansion is 
\begin{equation}
    \Psi = \sum_{|m|\leq N/2} a_m \phi_m,
\end{equation}
where the expansion coefficients satisfy $\sum_{|m|\leq N/2} |a_m|^2 = 1$. 
The discretized Fourier mode on the $j$-th grid point is $\phi_{m,j} =\frac{1}{\sqrt{N}} e^{2 \pi i m j / N}$, and $\phi_m$ is an eigenmode of the unitary shifting operator
$(\hat{U}^\pm \phi_m)_j = \phi_{m,j\mp1} = e^{\mp 2 \pi i m / N} \phi_{m,j}$.
Applying the advection operator $\hat{T}^\pm$ to $\Psi$ gives
\begin{equation}
    \hat{T}^\pm \Psi 
    = \sum_{|m|\leq N/2} a_m [(1-f) + f\hat{U}^\pm]\phi_m
    = \sum_{|m|\leq N/2} a_m G_m\phi_m,
\end{equation}
where $G_m = (1-f) + fe^{\mp 2 \pi i m / N}$ is the eigenvalue of $\phi_m$ under $\hat{T}^\pm$.
The norm is $|G_m|^{2} = 1 - F\sin^2(\pi m/N)$, where $F=4f(1-f)\in [0,1]$.
The norm $|G_m|^{2}\le 1$, and the equality is attained only when 
(i) $m=0$, which is consistent with the fact that a constant function is invariant under advection, and (ii) when $f=0$ or $1$, which is consistent with the fact that $\hat{T}^\pm$ is unitary only at these special $f$ values. 
After $S$ iterations, the final state is 
$(\hat{T}^\pm)^S \Psi = \sum_{|m|\leq N/2} a_m G^S_m\phi_m$.
The success probability is the norm $|(\hat{T}^\pm)^S \Psi|^2$, so $P_{\text{success}} = \sum_{|m|\leq N/2} |a_m|^2|G_m|^{2S}$.
The failure probability is $P_{\text{fail}} = 1- P_{\text{success}}$, which can be written as
\begin{equation}
    \label{eq:fail}
    P_{\text{fail}} = \sum_{|m|\leq N/2} |a_m|^2 (1-|G_m|^{2S}),
\end{equation}
For a desired advection distance $d/L$ at fixed $f$, the number of time steps $S$ is given by Eq.~(\ref{eq:advection_time}), which can be rewritten as $S=\gamma N$, 
where $\gamma = d/(Lf)$.
At a fixed $m$, when $N \to \infty$, we have $\pi m/N \to 0$, so 
\begin{align}
    |G_m|^{2S} 
    \simeq \left[1 - F\left(\pi m/N\right)^2\right]^{\gamma N} 
    \simeq 1 - \gamma F \pi^2 m^2/N,
    \label{Gm_to_1}
\end{align}
which means that the failure $1-|G_m|^{2S}$ goes to zero as $O(N^{-1})$ for fixed Fourier modes.
For any $\epsilon > 0$, because $\Psi \in L^2$, there exists some fixed $K$ such that $\sum_{|m| > K} |a_m|^2 < \epsilon$. We can split the total failure probability into two sums
\begin{align}
    P_{\text{fail}} = \Big(\sum_{|m|\leq K} + \!\sum_{K < |m|\leq N/2}\Big) |a_m|^2 (1-|G_m|^{2t})
    = O(N^{-1})+O(\epsilon).
\end{align}
Because $K$ is fixed for a given $\Psi$, the first sum goes to zero due to Eq. (\ref{Gm_to_1}) and $|a_m|^2 \le 1$.
Because $|G_m|^2 \in [0,1]$, the second sum is less than or equal to 
$\sum_{|m|>K} |a_m|^2 < \epsilon$.
Therefore, $P_{\text{fail}} < \epsilon$ for arbitrarily small $\epsilon$ when $N\to\infty$.
Although the above proof shows $P_{\text{success}} \to 1$ for all $L^2$ functions, the convergence can be arbitrarily slow. 
In Appendix B, we show that for smooth and piecewise smooth functions, $P_{\text{fail}} = O(N^{-1})$. For piecewise continuous functions with one or more discontinuities, $P_{\text{fail}} = O(N^{-1/2})$. More generally, if the amplitudes of the Fourier modes decay polynomially as $a_m = O(m^{-\alpha})$, we show that $P_{\text{fail}} = O(N^{1/2 - \alpha})$ for $1/2 < \alpha < 3/2$, and $P_{\text{fail}} = O(N^{-1})$ for $\alpha > 3/2$. For $\alpha \leq 1/2$, the function is no longer $L^2$, and $P_{\text{fail}}$ does not converge to zero.

\begin{figure}[H]
    \begin{center}
	\includegraphics[width=100mm]{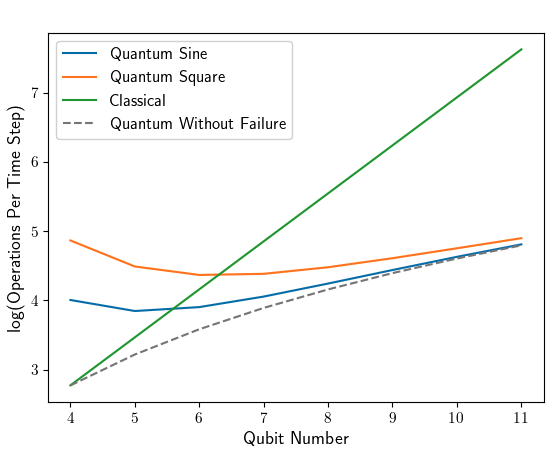}
    \end{center}
    \caption[Operations Per Time Step with Non-Zero Failure Probabilities]{
	The expected number of operations per time step for the classical algorithm (green) and the quantum algorithm. In the absence of failure, the complexity of the ideal quantum algorithm (dashed gray line) attains an exponential speedup over the classical algorithm. In the presence of failure, whose probability depends on the initial waveforms, the non-ideal complexity for the sine (blue) and the square (orange) waveform is higher [Eq.~(\ref{exact_complexity})]. Nevertheless, the non-ideal complexity converges to the ideal case when the number of qubits increases. The non-ideal complexity is evaluated for propagation across one box length and for the worst-case $f = 1/2$.
	}
\label{operations_per_step}
\end{figure}

For $L^2$ functions, because $P_{\text{success}}\to 1$, the average complexity per time step converges to the ideal complexity $O(n^2)$ when $n\to \infty$. 
We demonstrate the convergence empirically in Fig. \ref{operations_per_step}, where the initial waveforms are sine and square.
The failure probability is calculated using Eq.~(\ref{eq:fail}) for advection distance $d/L = 1$ and $f = 1/2$. 
Using $P_{\text{fail}}$ and Eq.~(\ref{exact_complexity}), the average complexity per time step is calculated. 
At high qubit numbers, the complexity approaches the ideal quantum complexity (dashed gray) of $O(n^2)$ per time step for both the sine (blue) and square (orange) waveforms. Meanwhile, the classical algorithm always has complexity $O(2^n)$. Thus, even with failures at each time step, our quantum advection algorithm still exhibits exponential speedup per time step over classical algorithms for large qubit numbers.

Finally, it is worth mentioning caveats to the above convergence arguments. In practice, at a finite number of qubits, the complexity per time step is always worse than $O(n^2)$. 
Using more qubits on noisy quantum hardware also increases failures due to hardware imperfections. 
Additionally, the time complexity of our quantum advection algorithm is the same as classical, namely $O(2^n)$, which will dominate the overall complexity at high qubit numbers. In practice, a balance must be found between minimizing the complexity per time step and minimizing the overall computational cost. The failure recovery algorithm is therefore appealing because it increases $P_{\text{success}}$, thus reducing the complexity without increasing qubit number.

\subsection{Failure Recovery Efficiency}
The failure recovery algorithm consists of quantum Fourier transform (QFT), the application of the matrix $\hat{K}$, inverse QFT, and a final half-step advection. Both QFT and inverse QFT have well-known quantum circuits \cite{coppersmith_approximate_2002} with asymptotic complexity $O(n^2)$. The final half-step advection has complexity $O(n^2)$ as discussed in Sec.~\ref{sec:advection_efficiency}. 
The final advection has an additional small probability of failure, in which case we perform a full restart, although in principle the failure recovery can be applied recursively.
In this section, we focus on the complexity of $\hat{K}$.

The gate complexity of an approximate $\hat{K}$ can be reduced to $O(n)$ by truncating the Fourier modes.
As shown in Sec.~\ref{sec:decompose_K}, the exact decomposition of $\hat{K}$ requires editing $2^n$ entries. Each edit requires $O(n)$ number of $X$ gates and a multi-controlled rotation, which takes no more than $16n-40$ controlled NOT gates \cite{vale_decomposition_2023}. 
Thus, the total gate complexity of $\hat{K}$ is $O(n2^n)$, which is exponentially worse than QFT and the advection per time step. 
To make the failure recovery algorithm useful, we approximate $\hat{K}$ by truncating the Fourier modes, such that $\hat{K}$ is not a bottleneck. 
For a waveform dominated by low Fourier modes, truncating $\hat{K}$ is a good approximation because the diagonal elements of $\hat{K}$ have magnitude $1/k$, which goes to zero as shown in Fig.~\ref{k_shape}. 
We edit $\hat{K}$ from a template created by applying an initial $X$ gate on the ancilla. The template is the $k\to\infty$ limit where the diagonal blocks are zero and the two off-diagonal blocks are identities. 
Combining these observations, we approximate $\hat{K}$ by only editing the first and last few elements of the diagonal blocks, while leaving the remaining diagonal entries in the middle as zeros. The off-diagonal blocks are edited or left unchanged accordingly.
Suppose $l$ is an integer between $0$ and $n-1$. We edit only the first and last $2^l$ entries of $\hat{K}$. When $l=0$, no integration is performed. On the other hand, when $l=n-1$, full integration is performed without approximation. Between these extremes, an approximate quantum indefinite integration is performed. The gate complexity of implementing the approximate $\hat{K}$ becomes $O(n2^{l+1})$. 
Truncating $\hat{K}$ effectively truncates the Fourier series when performing the indefinite integral. 
In our approximation, we only keep the first $2^l$ Fourier modes and their negative counterparts. 
Unless the initial waveform has extremely high-frequency modes, $l$ does not need to be large to achieve a reasonable approximation.
In practice, due to numerical diffusion, it is sufficient to set $l$ to $4$ or $5$, which corresponds to $16$ to $32$ Fourier modes, even for the square waveform. 
If we restrict $l$ to a small constant value, the asymptotic complexity for implementing the approximate $\hat{K}$ is therefore reduced to $O(n)$.

The failure recovery algorithm itself has a non-zero probability of failure, which depends strongly on the Fourier spectrum of the waveform and only weakly on the number of qubits. 
For successful integration, the ancilla must be measured as $\ket{0}$ after applying $\hat{K}$. Only then does the state of data qubits collapse into the Fourier transform of the desired integral. If the ancilla is measured as $\ket{1}$, the recovery fails. 
The success probability of the integration is proportional to $|\frac{1}{k} \text{QFT}(\Phi)|^2$, where $\Phi$ is the failure state. Therefore, success depends heavily on the spectrum of the function. When the spectrum is dominated by low Fourier modes, the success probability is high because $1/k$ is of order unity for small $k$. However, when the spectrum has significant contributions from high Fourier modes, because $1/k$ quickly decreases to zero, the success probability diminishes. Table~\ref{recovery_probs} shows the success probability of failure recovery for example functions. 
The probabilities are computed using classical emulations with $10^4$ shots and no truncation.  
Unlike the advection algorithm, the success probability of failure recovery does not depend significantly on the number of qubits. As long as the function is well-resolved, adding more qubits will not change the Fourier spectrum, therefore leaving the success probability unchanged. The only exceptions to this behavior are functions that require infinite Fourier modes like square waves. For these functions, adding more qubits results in higher order Fourier modes being occupied, reducing the amplitude of the lower order Fourier modes to maintain normalization. In this case, adding more qubits reduces the success probability of the failure recovery algorithm.

\begin{table}[htbp]
	\centering
	\renewcommand{\arraystretch}{1.6}
	\setlength{\extrarowheight}{3pt}	
	\begin{tabular}{cc}
		\hline
		Initial Function & Recovery Probability \\
		\hline
		$\displaystyle \sin\left(\frac{2\pi x}{L}\right)$
		& 1 \\
		$\displaystyle \sin\left(\frac{4\pi x}{L}\right)$
		& 0.25 \\
		$\displaystyle \sin\left(\frac{6\pi x}{L}\right)$
		& 0.11 \\
		$\displaystyle
		\sin\left(\frac{4\pi x}{L}\right)
		+
		\sin\left(\frac{6\pi x}{L}\right)$
		& 0.15 \\
		Square
		& 0.06 \\
		Linear $\displaystyle \left(\frac{x}{L}\right)$
		& 0.05 \\
		Quadratic $\displaystyle \left(\frac{x}{L}-\frac{1}{2}\right)^2$
		& 0.66 \\
		\hline
	\end{tabular}
	    \caption[Failure Recovery Probabilities]{
		Success probabilities of failure recovery for 6 data qubits. 
        An initial function is classically differentiated to prepare a failure state, which is then fed into quantum Fourier transform, followed by the application of $\hat{K}$, and a measurement on the ancilla. The failure recovery succeeds when the ancilla ends in the $\ket{0}$ state. 
        The probabilities are obtained using classical emulations with $10^4$ shots. 		
		The success probability is higher if the initial function is dominated by lower Fourier modes.
	}
	\label{recovery_probs}
\end{table}

\subsection{Combined Efficiency}
\begin{figure}[H]
    \begin{center}
	\includegraphics[width=\linewidth]{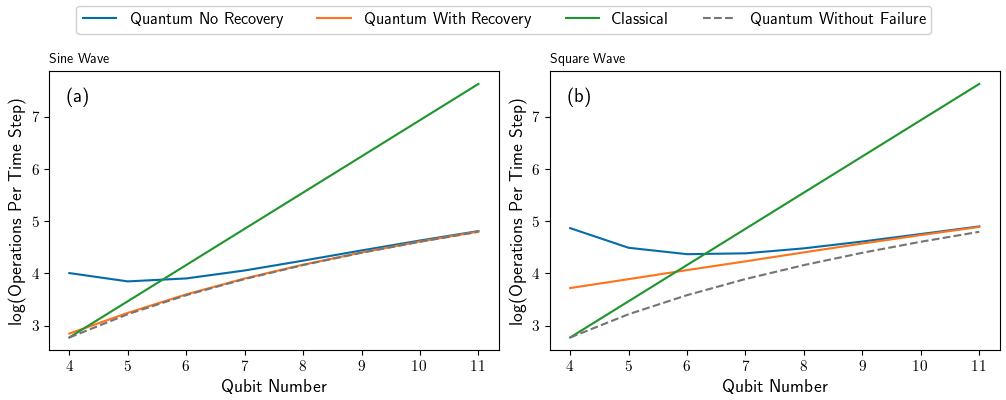}
    \end{center}
    \caption[Effect of Failure Recovery on Advection Complexity]{
	The expected number of operations per time step to reach some advection distance is reduced by failure recovery. The green line is the classical complexity and the gray dashed line is the complexity of ideal quantum advection with no failure. For both sine (a) and square (b) waveforms, applying failure recovery (orange) reduces the complexity compared to without recovery (blue), especially at low qubit numbers. The non-ideal quantum complexities are evaluated for propagation across one box length and for the worst-case $f = 1/2$.
	}
\label{recovery_effect}
\end{figure}

Using the approximate failure recovery, both the LCUs advection and the failure recovery algorithms have a complexity of $O(n^2)$ per time step in the large $n$ limit, achieving exponential speedup over classical algorithms.
In Table~\ref{recovery_probs}, the failure recovery probability is low for many functions, which means that the quantum indefinite integration algorithm is in general ineffective as a stand-alone algorithm. 
Nevertheless, it can meaningfully increase the overall efficiency of shock-capturing quantum advection. 
As advection proceeds, the success probability of failure recovery increases. This is because numerical diffusion inherent to the upwind scheme reduces higher order Fourier modes, moving the spectral weight towards lower order modes \cite{hirsch_numerical_1988}. Since the success probability is higher for lower Fourier modes, the overall efficiency of our algorithm increases as simulations progresses. 
In Fig.~\ref{recovery_effect}, we show the effect of failure recovery. 
At each time step, the probability of unrecoverable failure is equal to the probability of advection failure multiplied by the probability of recovery failure, including the additional half-step advection. The unrecoverable failure is used to calculate the overall $P_{\text{success}}$, which determines the complexity per step according to Eq.~(\ref{exact_complexity}). 
For an initial sine waveform [Fig.~\ref{recovery_effect}(a)], near-perfect recovery is attained due to its simple Fourier decomposition.
For an initial square waveform [Fig.~\ref{recovery_effect}(b)], which has infinite non-zero Fourier modes, applying failure recovery reduces the expected number of operations per time step needed for advecting a given distance. 
Because failure recovery decreases the complexity per time step without increasing the number of time steps needed to reach a given advection distance, it is especially useful in minimizing the total complexity at low qubit numbers that are feasible for near-term quantum hardware.

\section{Demonstration of Full Algorithm}
In Sec.~\ref{sec:demonstration_advection}, we demonstrate upwind quantum advection based on LCUs using a state-vector simulator. In this section, we demonstrate the stand-alone failure recovery algorithm, as well as the full advection algorithm with failure recovery. 
For each demonstration, we first emulate the quantum algorithms classically, and then demonstrate the algorithms on Quantinuum's trapped-ion quantum devices.

\subsection{Demonstration of Failure Recovery}
\begin{center}
\begin{figure}[H]
	\centering
  \includegraphics[width=0.8\linewidth]{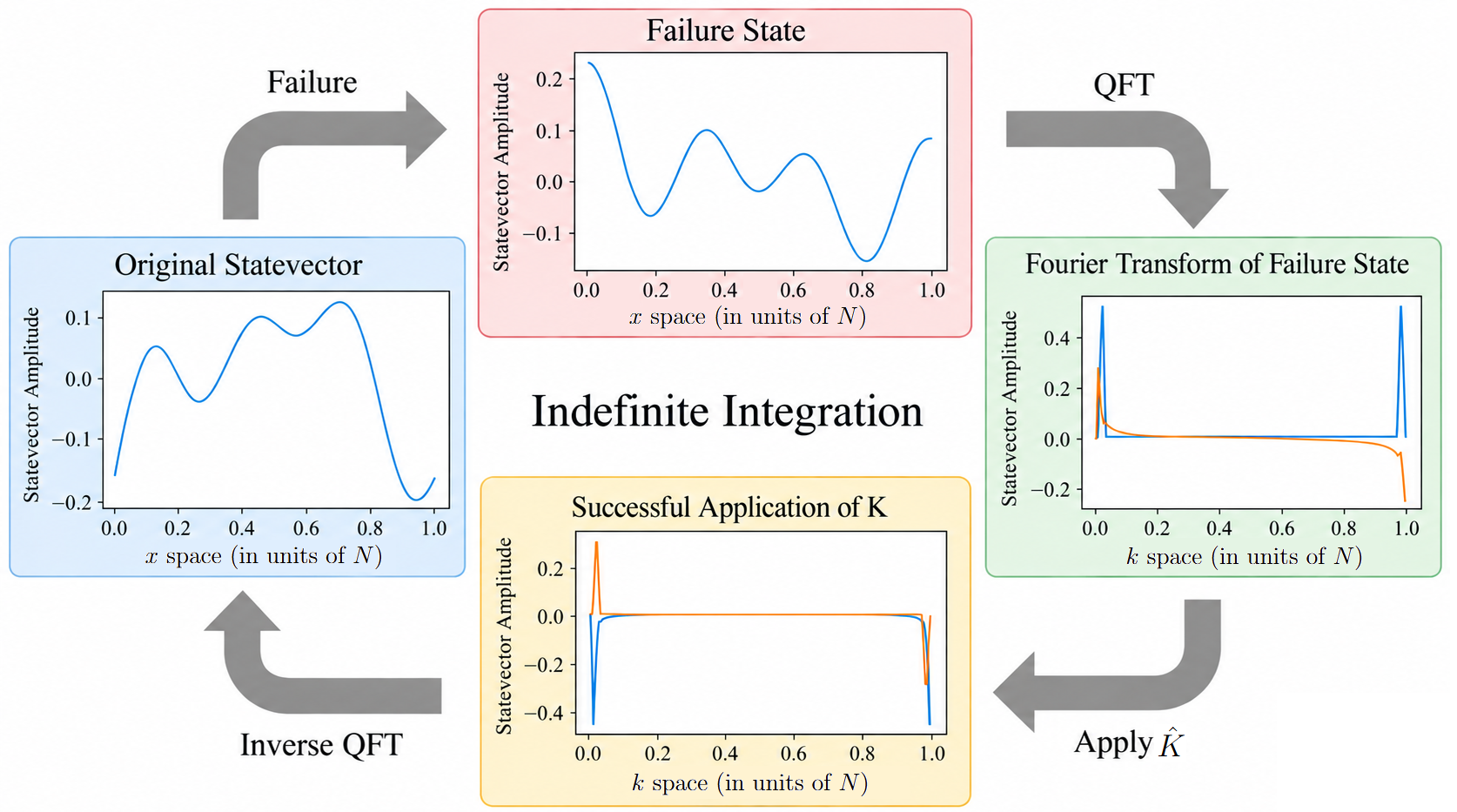}
  \caption{The failure recovery scheme, which achieves quantum indefinite integration, is demonstrated using Qiskit's state vector simulator for the example $\sin(\frac{4\pi x}{L}) + \sin(\frac{6\pi x}{L}) - 16(\frac{x}{L} - \frac{1}{2})^2$ with $7$ data qubits. 
  In Fourier space, blue and orange lines are the real and imaginary parts, respectively. The scheme takes a failure state, applies quantum Fourier transform (QFT), multiplies by $\frac{i}{k}$, and applies inverse QFT to recover the original state, up to an overall constant offset and normalization.}
  \label{integration_figure}
\end{figure}
\end{center}

As a stand-alone demonstration of the failure recovery algorithm, Figure \ref{integration_figure} shows a concrete example using a classical state-vector simulator. 
The left plot (blue) shows the initial state vector, which is given by $\Psi(x) = \sin(\frac{4\pi x}{L}) + \sin(\frac{6\pi x}{L}) - 16(\frac{x}{L} - \frac{1}{2})^2$. The spatial integral of this waveform is set to zero before runtime, as desired in Sec.~\ref{sec:quantun_integral}. 
The top plot (red) shows the failure state, resulting from a failed advection attempt with a positive advection velocity. 
The failure state $\Phi(x)=\Delta x \partial_x\Psi(x)$ is a numerical derivative of the initial state vector.
The right plot (green) shows the Fourier transform of the failure state, which is performed using QFT. In the Fourier space plots, the blue line is the real part, and the orange line is the imaginary part. 
The bottom plot (orange) shows the successful application of the $\hat{K}$ operator, where the ancilla is measured in the $\ket{0}$ state. 
Finally, inverse Fourier transform is applied, returning the state vector to the original state before failure, after performing a half-step advection.

\subsection{Demonstration of Advection with Failure Recovery}
\begin{figure}[H]
\centering
  \includegraphics[width=\linewidth]{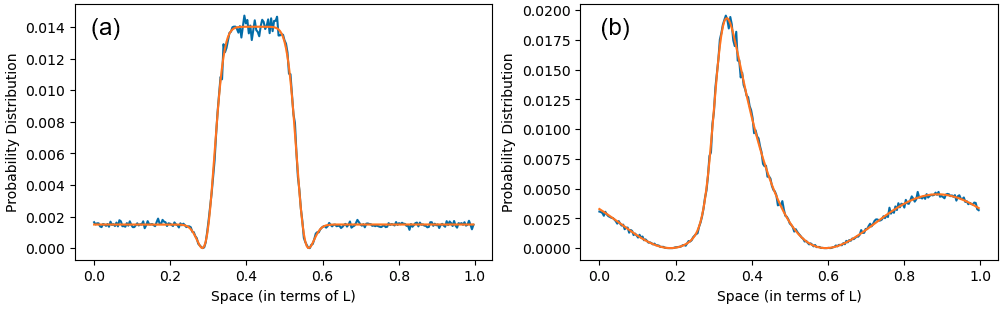}
  \caption[Full Advection Demonstration]{
	Advection algorithm with failure recovery is applied to (a) square waveform with a negative offset and (b) $\sin(\frac{2x}{L}) + \sin(\frac{3x}{L}) - (x - \frac{L}{2})^2$. For both examples, the spatial integral of the waveform is zero. 
    Using Qiskit's Aer simulator with dynamic circuit capabilities, the measured quantum probability distribution with $10^5$ shots (blue) matches the squared results of a classical upwind scheme applied to the same functions (orange). 
    The examples use $2^8$ grid points, $f=1/2$, and the advection distance is $d = 0.3L$. 
    The failure recovery is approximate, using an $l=4$ truncation. 
    The empirical overall success probability is $0.872$ for the square-like waveform and $0.957$ for the sine-like waveform.
	}
\label{advection_with_recovery}
\end{figure}

Figure \ref{advection_with_recovery} shows the results of our full advection algorithm with failure recovery using Qiskit's Aer simulator \cite{qiskit2024} for $n=8$ data qubits. 
In order to implement failure recovery, dynamic quantum circuits are used to allow mid-circuit measurements and conditional feedforward. 
When an advection step is successful, the next advection step is followed. 
On the other hand, if an advection step fails, the indefinite integration algorithm is performed with an $l=4$ truncation. If the failure recovery succeeds, an $f=1/2$ advection step is performed, returning the waveform to its state before the failed advection. From the recovered state, subsequent advection steps are performed. 
When the failure recovery itself fails, due to limitations of Qiskit's dynamic circuits, we do not perform a full restart. Instead, we simply discard the shot during post processing. This post-processing approach results in simpler circuits and shorter overall run times. 
In Fig.~\ref{advection_with_recovery}, the initial waveforms are prepared to have a zero spatial integral. As discussed in Sec.~\ref{sec:quantun_integral}, this constraint prevents the failure recovery algorithm from distorting the waveform.  
The probabilities shown in Fig.~\ref{advection_with_recovery} should be understood as the absolute value of the waveform, because the measured probability does not distinguish positive and negative waveforms.
To demonstrate the flexibility of our algorithm, we show results both for a square waveform [Fig.~\ref{advection_with_recovery}(a)] and a waveform that has quadratic and sine components [Fig.~\ref{advection_with_recovery}(b)]. 
For both initial waveforms, results of the quantum algorithm (blue) match classically expected results (orange), and the discrepancies are dominated by shot noise. 
Even with approximate failure recovery ($l=4$), the final probability distribution is hardly affected, supporting the validity of the truncation. These demonstrations show the accuracy of our quantum advection algorithm, which has exponential speedup over classical algorithms per time step.

\section{Discussion}
The algorithm presented in this paper solves the advection problem on quantum computers in one dimension with periodic boundary conditions. Our algorithm can capture shock-like discontinuities and has an exponential speedup over classical algorithms per time step. 
The ability to capture shocks comes from using a non-unitary upwind scheme instead of a unitary Hamiltonian approach that introduces spurious oscillations near discontinuities. 
Importantly, our non-unitary approach does not sacrifice exponential quantum speedup: For each advection step, the gate counts of our quantum circuits scale quadratically with the number of qubits in the large $n$ limit. Our algorithm can be extended to higher spatial dimensions by simply performing advection separately in each dimension. 

One question that naturally arises from any quantum solution to differential equations is how to extract information at the end of the simulation. Because each measurement can only give one bit string, there is no way to extract the full quantum state vector without performing the algorithm $O(2^n)$ times and measuring the result after each iteration. Such an approach would obliterate any speedup over classical algorithms \cite{aaronson_learnability_2007}. One way around this measurement problem is to apply specific operators to the final state, allowing us to extract partial information such as dominant frequency, local amplitudes, energy, and other physical observables. What observables are meaningful and how the observables can be measured is problem-dependent. Some possibilities include amplitude estimation \cite{lomonaco_quantum_2002}, dominant frequency analysis using QFT, and Pauli string expectation values \cite{peruzzo_variational_2014}.  
In this paper, we have only focused on how to enact the advection operator, which is often just one term in a more complicated problem. 

Our advection algorithm assumes periodic boundary conditions, which may not be suitable for certain use cases. This periodicity is built into the advection operators $\hat{T}^\pm$ and is required for the proposed LCUs decomposition. Therefore, it is unclear how this framework could be applied to other boundary conditions, which may present a significant limitation to its applicability. The algorithm remains to be extended to more general boundary conditions.

Unlike existing quantum advection solvers, our algorithm has a probability of failure at each time step. However, we prove that for enough qubits, the failure does not affect the algorithm's complexity per time step, therefore maintaining an exponential quantum speedup. This result must be balanced with the time complexity, which scales exponentially with qubit number due to the CFL condition. By adding qubits, one trades reduced time step complexity for an increased number of time steps.
To improve the success probability without increasing the time complexity, we develop a failure recovery algorithm, which is effectively an algorithm that performs quantum indefinite integration of an unknown quantum state. The algorithm is inefficient for exact integration, but has a truncated form that approximates integrals exponentially faster than classical algorithms. The integration algorithm has a high probability of failure, making it ineffective as a stand-alone tool. However, when used as a failure recovery algorithm, it reduces the complexity per advection time step without increasing the time complexity.

Our algorithm can be used as a subroutine in larger physics simulations, particularly plasma or fluid simulations with an advection term. Such a simulator could do an advection step and then a nonlinear step, alternating between the two to get an accurate and fast simulation of nonlinear dynamics using Hamiltonian splitting \cite{crouseilles_hamiltonian_2015}. The advection step can easily handle varying speeds, as long as $f\leq1$. With an exponential speedup for the advection term, plasma and fluid simulations may run more efficiently on quantum computers than what is possible on classical computers. 
Efficient quantum algorithms for solving the non-linear parts of the problems remain to be developed.




%

\section*{Acknowledgements}
This material is based upon work supported by the U.S. Department of Energy Office of Science 
under Award Number DE-SC0020393.
The access to Quantinuum is provided by the Quantum Computing User Program of the Oak Ridge Leadership Computing Facility at the Oak Ridge National Laboratory, which is supported by the Advanced Scientific Computing Research programs in the Office of Science of the U.S. Department of Energy under Contract No. DE-AC05-00OR22725.

\section*{Code and Data Availability}
The computer codes underlying results in this paper are openly available on GitLab at
\url{https://gitlab.com/seanYuanSHI/lattice-advection}.
The dataset underlying figures in this paper are openly available on Zenodo at
\url{https://zenodo.org/}.

\section*{Appendix}

\subsection*{A. Proof of $\hat{U}^+$ Gate Decomposition\label{App:Gate}}
We focus on the positive shifting operator $\hat{U}^+$. The negative shifting operator $\hat{U}^- = (\hat{U}^+)^\dagger$ is its Hermitian conjugate, which means the gate sequence for $\hat{U}^-$ is the reverse of that for $\hat{U}^+$.
For one data qubit ($n=1$), each entry should shift to the right by one grid point. 
Under periodic boundary conditions, the shift is $\ket{0} \leftrightarrow \ket{1}$. 
Thus, $\hat{U}^+ = \begin{bsmallmatrix} 0 & 1 \\ 1 & 0 \end{bsmallmatrix} = X$ is the Pauli $X$ gate, which matches the gate decomposition in Sec.~\ref{sec:U_decomposition} trivially.

Assume the proposed gate decomposition for $\hat{U}^+$ is true for $n=k$ qubits with $k \geq 1$. To continue the pattern for the $n=k+1$ case, we first add the $(k+1)$-th qubit as the most significant bit.
The $k$-qubit positive shifting operator $\hat{U}^+_k$ in the enlarged $2^{k+1}$ vector space is 
\begin{equation}
   \begin{bmatrix}
   0 &  & \dots & 0 & 1 & 0 &&\dots&& 0 \\
   1 & 0 &  & & 0 &&&&& \\
   0 & 1 & 0 & & \vdots & \vdots &&\ddots&&\vdots \\
   \vdots &  & \ddots & \ddots & 0 &&&&&\\
   0 & \dots & 0 & 1 & 0 & 0 &&\dots&& 0 \\
   0 &&\dots&& 0 & 0 & & \dots & 0 & 1 \\
   &&&&& 1 & 0 & & & 0 \\
   \vdots &&\ddots&&\vdots& 0 & 1 & 0 & & \vdots \\
   &&&&& \vdots & & \ddots & \ddots & 0 \\
   0 &&\dots&& 0 & 0 & \dots & 0 & 1 & 0 \\
   \end{bmatrix},
   \label{U_k_in_k+1_space}
\end{equation}
where the matrix is $2^{k+1} \times 2^{k+1}$ dimensional. 
Second, we apply a multi-controlled NOT gate with target qubit $k+1$ and control qubits $1$ through $k$ inclusive. 
The multi-controlled gate should map $\ket{0}\otimes\ket{1}^{\otimes k} \leftrightarrow \ket{1}\otimes\ket{1}^{\otimes k}$,
while leaving every other entry unchanged. Thus, the multi-controlled NOT gate can be written in matrix form as
\begin{equation}
\begin{bmatrix}
   1 & 0 & \dots & & 0 & 0 &&\dots&& 0 \\
   0 & 1 & & & \vdots &&\ddots&&& \\
   \vdots & & \ddots & & & \vdots &&&&\vdots \\
    &  & & 1 & 0 &&&& 0 & 0 \\
   0 & & \dots & 0 & 0 & 0 &&\dots& 0 & 1 \\
   0 &&\dots&& 0 & 1 & 0 & \dots & & 0 \\
   &\ddots&&& \vdots & 0 & 1 & & & \vdots \\
   \vdots &&&&& \vdots & & \ddots & & \\
   &&& 0 & 0 & & & & 1 & 0 \\
   0 &&\dots& 0 & 1 & 0 & & \dots & 0 & 0 \\
\end{bmatrix},
\label{multi-controlled-not_matrix}
\end{equation}
which is again a $2^{k+1} \times 2^{k+1}$ dimensional matrix.
Multiplying Eqs.~(\ref{U_k_in_k+1_space}) by (\ref{multi-controlled-not_matrix}), the overall effect of $\hat{U}^+_k$ followed by the multi-controlled NOT gate has the matrix representation
\begin{equation}
\begin{bmatrix}
    0 & & \dots & 0 & 1 \\
    1 & 0 & & & 0 \\
    0 & 1 & 0 & & \vdots \\
    \vdots &  & \ddots & \ddots & 0 \\
    0 & \dots & 0 & 1 & 0 \\
\end{bmatrix},
\end{equation}
which is precisely the $(2^{k+1} \times 2^{k+1})$ matrix $\hat{U}^+_{k+1}$. We have thus completed the proof by induction.

\subsection*{B. Proof of Failure Probability Scalings\label{App:scaling}}
For a desired advection distance $d$ in a box of length $L$, the advection algorithm has a total chance of failure $P_{\text{fail}}$, which is compounded over time steps. The probability of failure at each time step is equal to the squared norm of the upwind derivative of the waveform at that time step. Each time step, the advection operator $\hat{T}^\pm$ acts on the Fourier modes $\phi_m$ with eigenvalues $G_m = (1-f) + fe^{\mp 2 \pi i m / N}$, where $N$ is the number of spatial grid points and $m$ is the Fourier mode index.
For an initial state given by the Fourier series $\Psi = \sum_m a_m\phi_m$, the failure probability at time step $t$ is
\begin{equation}
    p_{\text{fail}}^{(t)} = \frac{f(1-f)}{\sum_{m\le N/2 }|a_m|^2 |G_m|^{2t}}\sum_{m\le N/2 } |D_m|^2 |a_m|^2 |G_m|^{2t},
\end{equation}
where $D_m = e^{- 2 \pi i m / N} - 1$ is the eigenvalue of the upwind finite difference operator in Fourier space. 
The $f(1-f)$ factor comes from LCUs, and the denominator $\sum_m |a_m|^2 |G_m|^{2t}$ is the normalization factor of the state prior to the step. 
From Sec.~\ref{sec:advection_efficiency}, this normalization constant converges to $1$ at large $N$ for all $L^2$ functions. 
Because the denominator is bounded from below, we drop it in the following estimations.
The eigenvalues satisfy inequalities
\begin{align}
    |D_m|^2 &= 4\sin^2(\pi m/N) \leq 4 (\pi m/N)^2, \\
    |G_m|^{2t} &= [1 - 4f(1-f)\sin^2(\pi m/N)]^t 
               \leq e^{-4tf(1-f)\sin^2(\pi m/N)t} 
               \leq e^{-4tf(1-f)(\pi m/N)^2},
\end{align}
where the last inequality holds because $|m| \leq N/2$.
We want to show the failure probability is bounded for $L^2$ functions, and the upper bound goes to zero when $N\to\infty$.

Assume $a_m$ scales as some inverse power of $m$ such that $|a_m|\le A |m|^{-\alpha}$, where $\alpha, A>0$ are constants that depend only on the initial function $\Psi$. Then, the probability of failure at time step $t$ as bounded by
\begin{equation}
    p_{\text{fail}}^{(t)} \leq f(1-f) \frac{4 \pi^2 A^2}{N^2} \sum_{m\le N/2} \frac{m^2}{m^{2\alpha}} e^{-4tf(1-f)(\pi m/N)^2},
    \label{p_fail_t_pre_cuttoff}
\end{equation}
which decreases with $N$ at higher resolution and decreases with $t$ due to numerical diffusion. 
This sum is a polynomial multiplied by a Gaussian of the form $Q_m = m^b e^{-a m^2}$, so large $m$ terms are suppressed and have negligible contribution to the sum after some characteristic cutoff $m>M$. 
The cutoff $M$ is on the order of either $a^{-1/2}$ or $M_c$, where $M_c$ is the $m$ value at which $Q_m$ attains its maximum. 
When $\alpha\ge 1$, we have $b\le 0$, so the maximum is attained at $M_c=0$. 
On the other hand, when $0<\alpha<1$, we have $b>0$ and the maximum is attained at $M_c=\sqrt{b/2a}$.
For both cases, we can estimate $M=O(N t^{-1/2})$.
Restricting the sum to this bound, replacing the sum by an integral, 
and grossly approximating $e^{-a m^2}=O(1)$, the integral becomes a polynomial and we find an upper bound $p_{\text{fail}}^{(t)} < O(\frac{1}{N^2}(\frac{N}{\sqrt{t}})^{3-2\alpha}).$
The algorithm fails if it fails at any step. An upper bound of the overall failure probability is 
\begin{align}
    P_{\text{fail}} < \sum_{t=0}^{S-1} p_{\text{fail}}^{(t)} < O\big(N^{1-2\alpha} (S - 1)^{\alpha - 1/2}\big),
\end{align}
where $S=Nd/(Lf) = \gamma N$ is the number of time steps needed to reach the desired distance. 
In the large $N$ limit, the above expression simplifies to 
$P_{\text{fail}} < O(N^{1/2-\alpha})$. 
This upper bound has three cases. 
(i) When $\alpha \leq 1/2$, for which $\Psi$ is not an $L^2$ function, the exponent of $N$ is positive, so the upper bound of $P_{\text{fail}}$ diverges. 
(ii) At $\alpha = 1/2$, which is the cutoff between $L^2$ and non-$L^2$ functions, the upper bound of $P_{\text{fail}}$ is $O(1)$.
(iii) When $\alpha>1/2$, the upper bound goes to zero, so $P_{\text{fail}}$ converges to zero when $N$ goes to infinity.

The convergence of $P_{\text{fail}}$ is stronger for larger $\alpha$. Using
Eq.~(\ref{p_fail_t_pre_cuttoff}), rather than $P_{\text{fail}} < O(N^{1/2-\alpha})$, we can arrive at similar conclusions with additional insights. 
(i) When $\alpha > 3/2$, the sum $\sum_m m^{2-2\alpha}$ converges to a finite value. 
In this case, numerical diffusion is not necessary for convergence, and the probability of failure at each time step scales as $O(N^{-2})$.
Summing over $S=\gamma N$ time steps, we see $P_{\text{\text{fail}}} < O(N^{-1})$.
A related case is for functions with a discrete number of Fourier modes, for which the sum $\sum_m m^2 |a_m|^2$ is bounded. This case is effectively $\alpha\to\infty$, so it also converges as $P_{\text{fail}} < O(N^{-1})$.
(ii) When $1/2 < \alpha < 3/2$, the sum $\sum_m m^{2-2\alpha}$ diverges, so diffusion is necessary for convergence and $P_{\text{fail}} < O(N^{1/2 - \alpha})$. 
(iii) At $\alpha = 3/2$, the sum $\sum_m m^{2-2\alpha} = O(\log N)$, so the probability of failure at each time step is bounded by $O(N^{-2} \log N)$. Summing over $S=\gamma N$ time steps, we see $P_{\text{fail}} < O(N^{-1}\log N)$, which is a looser upper bound. We see that numerical diffusion generally improves the convergence.





\pagebreak

\bibliographystyle{MSP}
\bibliography{Quantum_Advection_Project}

@misc{qiskit2024,
      title={Quantum computing with {Q}iskit},
      author={Javadi-Abhari, Ali and Treinish, Matthew and Krsulich, Kevin and Wood, Christopher J. and Lishman, Jake and Gacon, Julien and Martiel, Simon and Nation, Paul D. and Bishop, Lev S. and Cross, Andrew W. and Johnson, Blake R. and Gambetta, Jay M.},
      year={2024},
      doi={10.48550/arXiv.2405.08810},
      eprint={2405.08810},
      archivePrefix={arXiv},
      primaryClass={quant-ph}
}

@article{PhysRevA.110.062214,
  title = {Preparing angular momentum eigenstates using engineered quantum walks},
  author = {Shi, Yuan and Beck, Kristin M. and Kruse, Veronika Anneliese and Libby, Stephen B.},
  journal = {Phys. Rev. A},
  volume = {110},
  issue = {6},
  pages = {062214},
  numpages = {25},
  year = {2024},
  month = {Dec},
  publisher = {American Physical Society},
  doi = {10.1103/PhysRevA.110.062214},
  url = {https://link.aps.org/doi/10.1103/PhysRevA.110.062214}
}

@article{Courant52,
author = {Courant, Richard and Isaacson, Eugene and Rees, Mina},
title = {On the solution of nonlinear hyperbolic differential equations by finite differences},
journal = {Communications on Pure and Applied Mathematics},
volume = {5},
number = {3},
pages = {243-255},
doi = {https://doi.org/10.1002/cpa.3160050303},
url = {https://onlinelibrary.wiley.com/doi/abs/10.1002/cpa.3160050303},
eprint = {https://onlinelibrary.wiley.com/doi/pdf/10.1002/cpa.3160050303},
year = {1952}
}

@article{NOVIKAU2026115132,
title = {An efficient explicit implementation of a near-optimal quantum algorithm for simulating linear dissipative differential equations},
journal = {Journal of Computational Physics},
volume = {564},
pages = {115132},
year = {2026},
issn = {0021-9991},
doi = {https://doi.org/10.1016/j.jcp.2026.115132},
url = {https://www.sciencedirect.com/science/article/pii/S0021999126004845},
author = {I. Novikau and I. Joseph}
}

@article{NOVIKAU2025109498,
title = {Quantum algorithm for the advection-diffusion equation and the Koopman-von Neumann approach to nonlinear dynamical systems},
journal = {Computer Physics Communications},
volume = {309},
pages = {109498},
year = {2025},
issn = {0010-4655},
doi = {https://doi.org/10.1016/j.cpc.2025.109498},
url = {https://www.sciencedirect.com/science/article/pii/S0010465525000013},
author = {I. Novikau and I. Joseph}
}

@article{May2025second,
  title={Second quantization of nonlinear Vlasov-Poisson system for quantum computation},
  author={May, Michael Q and Qin, Hong},
  journal={arXiv:2506.01895},
  year={2025}
}

@article{Sundar26,
  title = {Simulating plasma wave propagation on a superconducting quantum chip},
  author = {Sundar, Bhuvanesh and Evert, Bram and Geyko, Vasily and Patterson, Andrew and Joseph, Ilon and Shi, Yuan},
  journal = {Phys. Rev. Appl.},
  volume = {25},
  issue = {2},
  pages = {024077},
  numpages = {24},
  year = {2026},
  month = {Feb},
  publisher = {American Physical Society},
  doi = {10.1103/lxr6-t7vb},
  url = {https://link.aps.org/doi/10.1103/lxr6-t7vb}
}

@article{Shi24, 
title={Simulating nonlinear optical processes on a superconducting quantum device}, 
volume={90}, 
DOI={10.1017/S0022377824001326}, 
number={6}, 
journal={Journal of Plasma Physics}, 
author={Shi, Yuan and Evert, Bram and Brown, Amy F. and Tripathi, Vinay and Sete, Eyob A. and Geyko, Vasily and Cho, Yujin and DuBois, Jonathan L. and Lidar, Daniel and Joseph, Ilon and et al.}, 
year={2024}, 
pages={805900602}
}

@article{Shi21,
  title = {Simulating non-native cubic interactions on noisy quantum machines},
  author = {Shi, Yuan and Castelli, Alessandro R. and Wu, Xian and Joseph, Ilon and Geyko, Vasily and Graziani, Frank R. and Libby, Stephen B. and Parker, Jeffrey B. and Rosen, Yaniv J. and Martinez, Luis A. and DuBois, Jonathan L.},
  journal = {Phys. Rev. A},
  volume = {103},
  issue = {6},
  pages = {062608},
  numpages = {8},
  year = {2021},
  month = {Jun},
  publisher = {American Physical Society},
  doi = {10.1103/PhysRevA.103.062608},
  url = {https://link.aps.org/doi/10.1103/PhysRevA.103.062608}
}

@article{Kyriienko21,
  title = {Solving nonlinear differential equations with differentiable quantum circuits},
  author = {Kyriienko, Oleksandr and Paine, Annie E. and Elfving, Vincent E.},
  journal = {Phys. Rev. A},
  volume = {103},
  issue = {5},
  pages = {052416},
  numpages = {22},
  year = {2021},
  month = {May},
  publisher = {American Physical Society},
  doi = {10.1103/PhysRevA.103.052416},
  url = {https://link.aps.org/doi/10.1103/PhysRevA.103.052416}
}

@article{Andress25, 
title={Quantum simulation of nonlinear dynamical systems using repeated measurement}, 
volume={91}, DOI={10.1017/S0022377825000091}, 
number={2}, 
journal={Journal of Plasma Physics}, 
author={Andress, Joseph and Engel, Alexander and Shi, Yuan and Parker, Scott}, 
year={2025}, 
pages={E55}
}

@article{Ye24,
    author = {Ye, Chuang-Chao and An, Ning-Bo and Ma, Teng-Yang and Dou, Meng-Han and Bai, Wen and Sun, De-Jun and Chen, Zhao-Yun and Guo, Guo-Ping},
    title = {A hybrid quantum-classical framework for computational fluid dynamics},
    journal = {Physics of Fluids},
    volume = {36},
    number = {12},
    pages = {127111},
    year = {2024},
    month = {12},
    issn = {1070-6631},
    doi = {10.1063/5.0238193},
    url = {https://doi.org/10.1063/5.0238193},
    eprint = {https://pubs.aip.org/aip/pof/article-pdf/doi/10.1063/5.0238193/20279143/127111_1_5.0238193.pdf},
}

@article{Tennie23,
      author = "F. Tennie and T. N. Palmer",
      title = "Quantum Computers for Weather and Climate Prediction: The Good, the Bad, and the Noisy",
      journal = "Bulletin of the American Meteorological Society",
      year = "2023",
      publisher = "American Meteorological Society",
      address = "Boston MA, USA",
      volume = "104",
      number = "2",
      doi = "10.1175/BAMS-D-22-0031.1",
      pages=      "E488 - E500",
      url = "https://journals.ametsoc.org/view/journals/bams/104/2/BAMS-D-22-0031.1.xml"
}

@article{Joseph23,
    author = {Joseph, I. and Shi, Y. and Porter, M. D. and Castelli, A. R. and Geyko, V. I. and Graziani, F. R. and Libby, S. B. and DuBois, J. L.},
    title = {Quantum computing for fusion energy science applications},
    journal = {Physics of Plasmas},
    volume = {30},
    number = {1},
    pages = {010501},
    year = {2023},
    month = {01},
    issn = {1070-664X},
    doi = {10.1063/5.0123765},
    url = {https://doi.org/10.1063/5.0123765},
    eprint = {https://pubs.aip.org/aip/pop/article-pdf/doi/10.1063/5.0123765/19821810/010501_1_online.pdf},
}

@article{Gonzalez-Conde25,
  title = {Quantum Carleman linearization efficiency in nonlinear fluid dynamics},
  author = {Gonzalez-Conde, Javier and Lewis, Dylan and Bharadwaj, Sachin S. and Sanz, Mikel},
  journal = {Phys. Rev. Res.},
  volume = {7},
  issue = {2},
  pages = {023254},
  numpages = {16},
  year = {2025},
  month = {Jun},
  publisher = {American Physical Society},
  doi = {10.1103/PhysRevResearch.7.023254},
  url = {https://link.aps.org/doi/10.1103/PhysRevResearch.7.023254}
}

@article{Joseph_2023,
doi = {10.1088/1751-8121/ad0533},
url = {https://doi.org/10.1088/1751-8121/ad0533},
year = {2023},
month = {nov},
publisher = {IOP Publishing},
volume = {56},
number = {48},
pages = {484001},
author = {Joseph, Ilon},
title = {Semiclassical theory and the Koopman-van Hove equation*},
journal = {Journal of Physics A: Mathematical and Theoretical}
}

@article{berry_quantum_2017,
	title = {Quantum {Algorithm} for {Linear} {Differential} {Equations} with {Exponentially} {Improved} {Dependence} on {Precision}},
	volume = {356},
	issn = {1432-0916},
	url = {https://doi.org/10.1007/s00220-017-3002-y},
	doi = {10.1007/s00220-017-3002-y},
	language = {en},
	number = {3},
	urldate = {2026-09-14},
	journal = {Communications in Mathematical Physics},
	author = {Berry, Dominic W. and Childs, Andrew M. and Ostrander, Aaron and Wang, Guoming},
	month = dec,
	year = {2017},
	pages = {1057--1081},
}

@article{childs_quantum_2020,
	title = {Quantum {Spectral} {Methods} for {Differential} {Equations}},
	volume = {375},
	issn = {1432-0916},
	url = {https://doi.org/10.1007/s00220-020-03699-z},
	doi = {10.1007/s00220-020-03699-z},
	language = {en},
	number = {2},
	urldate = {2026-09-14},
	journal = {Communications in Mathematical Physics},
	author = {Childs, Andrew M. and Liu, Jin-Peng},
	month = apr,
	year = {2020},
	pages = {1427--1457},
}

@article{childs_high-precision_2021,
	title = {High-precision quantum algorithms for partial differential equations},
	volume = {5},
	url = {https://quantum-journal.org/papers/q-2021-11-10-574/},
	doi = {10.22331/q-2021-11-10-574},
	language = {en-GB},
	urldate = {2026-09-14},
	journal = {Quantum},
	author = {Childs, Andrew M. and Liu, Jin-Peng and Ostrander, Aaron},
	month = nov,
	year = {2021},
	note = {Publisher: Verein zur Förderung des Open Access Publizierens in den Quantenwissenschaften},
	pages = {574},
}

@article{engel_linear_2021,
	title = {Linear embedding of nonlinear dynamical systems and prospects for efficient quantum algorithms},
	volume = {28},
	issn = {1070-664X},
	url = {https://doi.org/10.1063/5.0040313},
	doi = {10.1063/5.0040313},
	number = {6},
	urldate = {2026-09-14},
	journal = {Physics of Plasmas},
	author = {Engel, Alexander and Smith, Graeme and Parker, Scott E.},
	month = jun,
	year = {2021},
	pages = {062305},
}

@article{liu_efficient_2021,
	title = {Efficient quantum algorithm for dissipative nonlinear differential equations},
	volume = {118},
	url = {https://www.pnas.org/doi/10.1073/pnas.2026805118},
	doi = {10.1073/pnas.2026805118},
	number = {35},
	urldate = {2026-09-14},
	journal = {Proceedings of the National Academy of Sciences},
	author = {Liu, Jin-Peng and Kolden, Herman Oie and Krovi, Hari K. and Loureiro, Nuno F. and Trivisa, Konstantina and Childs, Andrew M.},
	month = aug,
	year = {2021},
	note = {Publisher: Proceedings of the National Academy of Sciences},
	pages = {e2026805118},
}

@article{wu_quantum_2025,
	title = {Quantum {Algorithms} for {Nonlinear} {Dynamics}: {Revisiting} {Carleman} {Linearization} with {No} {Dissipative} {Conditions}},
	volume = {47},
	issn = {1064-8275},
	shorttitle = {Quantum {Algorithms} for {Nonlinear} {Dynamics}},
	url = {https://epubs.siam.org/doi/10.1137/24M1665799},
	doi = {10.1137/24M1665799},
	number = {2},
	urldate = {2026-09-14},
	journal = {SIAM Journal on Scientific Computing},
	author = {Wu, Hsuan-Cheng and Wang, Jingyao and Li, Xiantao},
	month = apr,
	year = {2025},
	note = {Publisher: Society for Industrial and Applied Mathematics},
	pages = {A943--A970},
}

@article{gaitan_finding_2020,
	title = {Finding flows of a {Navier}–{Stokes} fluid through quantum computing},
	volume = {6},
	doi = {10.1038/s41534-020-00291-0},
	journal = {npj Quantum Information},
	author = {Gaitan, Frank},
	month = jul,
	year = {2020},
	pages = {61},
}

@article{higuchi_quantum_2025,
	title = {Quantum calculation for two-stream instability and advection test of {Vlasov}–{Maxwell} equations: numerical evaluation of {Hamiltonian} simulation},
	volume = {91},
	issn = {0022-3778, 1469-7807},
	shorttitle = {Quantum calculation for two-stream instability and advection test of {Vlasov}–{Maxwell} equations},
	url = {https://www.cambridge.org/core/journals/journal-of-plasma-physics/article/quantum-calculation-for-twostream-instability-and-advection-test-of-vlasovmaxwell-equations-numerical-evaluation-of-hamiltonian-simulation/103D3AAF4402670B5D260E6DF5869D89},
	doi = {10.1017/S0022377825100500},
	language = {en},
	number = {4},
	urldate = {2026-09-14},
	journal = {Journal of Plasma Physics},
	author = {Higuchi, Hayato and Pedersen, Juan William and Toyoizumi, Kiichiro and Yoshikawa, Kohji and Kiumi, Chusei and Yoshikawa, Akimasa},
	month = aug,
	year = {2025},
	pages = {E116},
}

@article{crouseilles_hamiltonian_2015,
	title = {Hamiltonian splitting for the {Vlasov}–{Maxwell} equations},
	volume = {283},
	issn = {0021-9991},
	url = {https://www.sciencedirect.com/science/article/pii/S0021999114007918},
	doi = {10.1016/j.jcp.2014.11.029},
	urldate = {2026-09-14},
	journal = {Journal of Computational Physics},
	author = {Crouseilles, Nicolas and Einkemmer, Lukas and Faou, Erwan},
	month = feb,
	year = {2015},
	pages = {224--240},
}

@article{over_quantum_2025,
	title = {Quantum algorithm for the advection-diffusion equation by direct block encoding of the time-marching operator},
	volume = {112},
	url = {https://link.aps.org/doi/10.1103/d8hb-fv93},
	doi = {10.1103/d8hb-fv93},
	number = {1},
	urldate = {2026-09-14},
	journal = {Physical Review A},
	author = {Over, Paul and Bengoechea, Sergio and Brearley, Peter and Laizet, Sylvain and Rung, Thomas},
	month = jul,
	year = {2025},
	note = {Publisher: American Physical Society},
	pages = {L010401},
}

@article{brearley_quantum_2024,
	title = {Quantum algorithm for solving the advection equation using {Hamiltonian} simulation},
	volume = {110},
	url = {https://link.aps.org/doi/10.1103/PhysRevA.110.012430},
	doi = {10.1103/PhysRevA.110.012430},
	number = {1},
	urldate = {2026-09-14},
	journal = {Physical Review A},
	author = {Brearley, Peter and Laizet, Sylvain},
	month = jul,
	year = {2024},
	note = {Publisher: American Physical Society},
	pages = {012430},
}

@article{hurricane_physics_2023,
	title = {Physics principles of inertial confinement fusion and {U}.{S}. program overview},
	volume = {95},
	url = {https://link.aps.org/doi/10.1103/RevModPhys.95.025005},
	doi = {10.1103/RevModPhys.95.025005},
	number = {2},
	urldate = {2026-09-14},
	journal = {Reviews of Modern Physics},
	author = {Hurricane, O. A. and Patel, P. K. and Betti, R. and Froula, D. H. and Regan, S. P. and Slutz, S. A. and Gomez, M. R. and Sweeney, M. A.},
	month = jun,
	year = {2023},
	note = {Publisher: American Physical Society},
	pages = {025005},
}

@article{barenco_elementary_1995,
	title = {Elementary gates for quantum computation},
	volume = {52},
	doi = {10.1103/PhysRevA.52.3457},
	number = {5},
	journal = {Physical Review A},
	author = {Barenco, Adriano},
	year = {1995},
	pages = {3457--3467},
}

@book{courant_partial_1956,
	title = {On the partial difference equations of mathematical physics},
	url = {http://archive.org/details/onpartialdiffere00cour},
	language = {eng},
	urldate = {2025-10-25},
	publisher = {New York: Courant Institute of Mathematical Sciences, New York University},
	author = {Courant, Richard and Lewy, H. and Courant, Richard and Friedrichs, Kurt Otto},
	collaborator = {New York University, Institute of Fine Arts Library},
	year = {1956},
}

@misc{coppersmith_approximate_2002,
	title = {An approximate {Fourier} transform useful in quantum factoring},
	url = {http://arxiv.org/abs/quant-ph/0201067},
	doi = {10.48550/arXiv.quant-ph/0201067},
	urldate = {2025-12-10},
	publisher = {arXiv},
	author = {Coppersmith, D.},
	month = jan,
	year = {2002},
	note = {arXiv:quant-ph/0201067},
}

@misc{vale_decomposition_2023,
	title = {Decomposition of {Multi}-controlled {Special} {Unitary} {Single}-{Qubit} {Gates}},
	url = {http://arxiv.org/abs/2302.06377},
	doi = {10.48550/arXiv.2302.06377},
	urldate = {2025-12-10},
	publisher = {arXiv},
	author = {Vale, Rafaella and Azevedo, Thiago Melo D. and Araújo, Ismael C. S. and Araujo, Israel F. and Silva, Adenilton J. da},
	month = feb,
	year = {2023},
	note = {arXiv:2302.06377 [quant-ph]},
}

@article{childs_hamiltonian_2012,
	title = {Hamiltonian simulation using linear combinations of unitary operations},
	volume = {12},
	issn = {1533-7146},
	number = {11-12},
	journal = {Quantum Info. Comput.},
	author = {Childs, Andrew M. and Wiebe, Nathan},
	month = nov,
	year = {2012},
	pages = {901--924},
}

@article{low_hamiltonian_2019,
	title = {Hamiltonian {Simulation} by {Qubitization}},
	volume = {3},
	url = {https://quantum-journal.org/papers/q-2019-07-12-163/},
	doi = {10.22331/q-2019-07-12-163},
	language = {en-GB},
	urldate = {2026-03-06},
	journal = {Quantum},
	author = {Low, Guang Hao and Chuang, Isaac L.},
	month = jul,
	year = {2019},
	note = {Publisher: Verein zur Förderung des Open Access Publizierens in den Quantenwissenschaften},
	pages = {163},
}

@article{aaronson_learnability_2007,
	title = {The learnability of quantum states},
	volume = {463},
	copyright = {https://royalsociety.org/journals/ethics-policies/data-sharing-mining/},
	issn = {1364-5021, 1471-2946},
	url = {https://royalsocietypublishing.org/doi/10.1098/rspa.2007.0113},
	doi = {10.1098/rspa.2007.0113},
	language = {en},
	number = {2088},
	urldate = {2026-03-08},
	journal = {Proceedings of the Royal Society A: Mathematical, Physical and Engineering Sciences},
	author = {Aaronson, Scott},
	month = dec,
	year = {2007},
	pages = {3089--3114},
}

@article{peruzzo_variational_2014,
	title = {A variational eigenvalue solver on a photonic quantum processor},
	volume = {5},
	copyright = {2014 The Author(s)},
	issn = {2041-1723},
	url = {https://www.nature.com/articles/ncomms5213},
	doi = {10.1038/ncomms5213},
	language = {en},
	number = {1},
	urldate = {2026-03-08},
	journal = {Nature Communications},
	author = {Peruzzo, Alberto and McClean, Jarrod and Shadbolt, Peter and Yung, Man-Hong and Zhou, Xiao-Qi and Love, Peter J. and Aspuru-Guzik, Alán and O’Brien, Jeremy L.},
	month = jul,
	year = {2014},
	note = {Publisher: Nature Publishing Group},
	pages = {4213},
}

@incollection{lomonaco_quantum_2002,
	address = {Providence, Rhode Island},
	title = {Quantum amplitude amplification and estimation},
	volume = {305},
	isbn = {978-0-8218-2140-4 978-0-8218-7895-8},
	url = {http://www.ams.org/conm/305/},
	language = {en},
	urldate = {2026-03-08},
	booktitle = {Contemporary {Mathematics}},
	publisher = {American Mathematical Society},
	author = {Brassard, Gilles and Høyer, Peter and Mosca, Michele and Tapp, Alain},
	editor = {Lomonaco, Samuel J. and Brandt, Howard E.},
	year = {2002},
	doi = {10.1090/conm/305/05215},
	pages = {53--74},
}

@book{leveque_finite_nodate,
	title = {Finite {Volume} {Methods} for {Hyperbolic} {Problems}},
	language = {en},
	author = {Leveque, Randall J},
    year = {2004},
    publisher = {Cambridge University Press},
}

@article{mottonen_transformation_2005,
	title = {Transformation of quantum states using uniformly controlled rotations},
	volume = {5},
	issn = {1533-7146},
	number = {6},
	journal = {Quantum Information \& Computation},
	author = {Möttönen, Mikko and Vartiainen, Juha J. and Bergholm, Ville and Salomaa, Martti M.},
	month = sep,
	year = {2005},
	pages = {467--473},
}

@book{hirsch_numerical_1988,
	address = {USA},
	title = {Numerical computation of internal \& external flows: fundamentals of numerical discretization},
	isbn = {978-0-471-91762-5},
	shorttitle = {Numerical computation of internal \& external flows},
	publisher = {John Wiley \& Sons, Inc.},
	editor = {Hirsch, Charles},
	month = jun,
	year = {1988},
}

@article{grzesiak_efficient_2020,
	title = {Efficient arbitrary simultaneously entangling gates on a trapped-ion quantum computer},
	volume = {11},
	issn = {2041-1723},
	doi = {10.1038/s41467-020-16790-9},
	language = {eng},
	number = {1},
	journal = {Nature Communications},
	author = {Grzesiak, Nikodem and Blümel, Reinhold and Wright, Kenneth and Beck, Kristin M. and Pisenti, Neal C. and Li, Ming and Chaplin, Vandiver and Amini, Jason M. and Debnath, Shantanu and Chen, Jwo-Sy and Nam, Yunseong},
	month = jun,
	year = {2020},
	pmcid = {PMC7289877},
	pmid = {32528164},
	pages = {2963},
}



\pagebreak

\end{document}